\documentclass[twocolumn,superscriptaddress,preprintnumbers,amsmath,amssymb,showpacs,floatfix]{revtex4-1}
\usepackage{graphicx}
\usepackage[export]{adjustbox}% http://ctan.org/pkg/adjustbox
\usepackage{dcolumn}% Align table columns on decimal point
\usepackage{amsmath,amsbsy,amssymb,scalefnt}
\usepackage{bm}% bold math
\usepackage{epstopdf}
\usepackage[T1]{fontenc}
\usepackage{textcomp}
\usepackage{mathtools}
\usepackage{breakcites}
\usepackage{hyperref}
\usepackage[mathscr]{eucal}
\usepackage{color,soul}
\DeclareMathAlphabet{\mathpzc}{OT1}{pzc}{m}{it}
\newcommand* {\koeff}[3]{#1^{#2}_{#3}}
\newcommand* {\vekc}[1]{{\ensuremath{\bm{\mathscr{#1}}}}}
\hypersetup{backref=true,
 pdfnewwindow=true, colorlinks=true,
 linkcolor=blue, anchorcolor=blue,
 citecolor=blue, filecolor=blue,
 menucolor=blue, urlcolor=blue}
\begin{document}

\title{\bf {Al$_2$B$_2$ and AlB$_4$ monolayers: emergence of multiple two-dimensional Dirac nodal line semimetals with novel properties}}
\date{\today} 
\author{Saeid Abedi}
\affiliation{Department of Physics, Isfahan University of Technology, Isfahan, 84156-83111, Iran}
\author{Esmaeil Taghizadeh Sisakht}\thanks{taghizadeh.sisakht@gmail.com}
\affiliation{Department of Physics, Isfahan University of Technology, Isfahan, 84156-83111, Iran}
\author{S. Javad Hashemifar}
\affiliation{Department of Physics, Isfahan University of Technology, Isfahan, 84156-83111, Iran}
\author{Nima Ghafari Cherati}
\affiliation{Department of Physics, Isfahan University of Technology, Isfahan, 84156-83111, Iran}
\author{Ismaeil Abdolhosseini Sarsari}
\affiliation{Department of Physics, Isfahan University of Technology, Isfahan, 84156-83111, Iran}
\author{Francois M. Peeters}
\affiliation{Department of Physics, University of Antwerp, Groenenborgerlaan 171, B-2020 Antwerpen, Belgium}

\begin{abstract}
%Two-dimensional (2D) materials with topological semimetal features have gained widespread interest due to their potential applications in developing nanoscale devices.
Topological semimetal phases in two-dimensional (2D) materials have gained widespread interest
due to their potential applications in developing nanoscale devices. 
Despite the prediction of the Dirac/Weyl points in a wide variety of 2D candidates, 
materials featuring topological nodal lines are still in great scarcity. 
Herein, we predict two stable thinnest films of aluminum diboride with 
hyper- and hypo-stoichiometries of Al$_2$B$_2$ and AlB$_4$ as new 2D
nonmagnetic Dirac nodal line semimetals (NLSMs) which promise to offer many novel features.
Our elaborate electronic structure calculations combined with analytical studies
reveal that, in addition to the multiple Dirac points,
these 2D configurations host various type-I closed nodal lines (NLs) around the Fermi level,
all of which are semimetal states protected by the time-reversal and in-plane mirror symmetries. 
The most intriguing NL in Al$_2$B$_2$ encloses the K point and crosses the Fermi level with a considerable dispersion, 
thus providing a fresh playground to explore exotic properties in dispersive Dirac nodal lines. 
More strikingly, in the case of 2D superconductor AlB$_4$ which exhibits a high 
transition temperature, 
%of about 47\,K [\cite{}],
we provide the first evidence for a set of 2D nonmagnetic open type-II NLs  in weak spin-orbit coupling limit,
coinciding with closed type-I NLs near the Fermi level.
The coexistence of superconductivity and nontrivial band topology in AlB$_4$
not only makes it a promising  material to  exhibit novel topological superconducting phases,
but also the rather large energy dispersion of type-II nodal lines in this configuration,
may offer a distinguished platform for realization of novel topological features in two-dimensional limit.
\end{abstract}

\maketitle
\footnotetext{\textit{$^{a}$~Address, Address, Town, Country. Fax: XX XXXX XXXX; Tel: XX XXXX XXXX; E-mail: xxxx@aaa.bbb.ccc}}
%\section{INTRODUCTION}
Topological classification of quantum states of matter has provided a new paradigm 
for the study of modern condensed matter physics. This vibrant area of research has emerged thanks to the 
discovery of two- (2D) and three-dimensional (3D) topological insulators 
(TIs)~\cite{bernevigQSH,konig2007quantum,fu2007topological1,moore2010birth,wang2013prediction,Zhao2011helical}.
TIs are insulating in bulk, while host conducting edge/surface states at the boundary that are protected by 
the time-reversal (TR) symmetry. 
Beyond the topological insulating phase, a new class of topological materials is realized
that exhibit a gapless bulk band structure and thus called topological semimetals (TSMs) or topological 
metals~\cite{Burkov2011Topological,Burkov2016Topological,Armitage2018Weyl,Yang2014Classification,Baik2015emergence}.\\
TSMs are characterized by the key features of the band crossing originated from the crystal symmetries. 
Depending on the type of band degeneracy, the codimension, and the band dispersion, 
one can categorize TSMs into three general classes of nodal Dirac\texttt{\small/}Weyl 
points~\cite{Wan2011Topological,Young2012Dirac,Liu2014Discovery,Soluyanov2015TypeII,Lv2015Observation,Xu2015Discovery,Chen2015nano}, 
nodal lines (NLs)~\cite{Kim2015Dirac,Mullen2015Line,Weng2015Topological,Hirayama2017Topological,Bzdusek2016Nodal,Zhang2017Coexistence,Feng2018Topological,
Liu2018Experimental,Tian2020creation},
and nodal surfaces~\cite{Liang085427,Zhong7232,Wu115125,Xu205310,Topp041073}.
The first class includes Dirac\texttt{\small/}Weyl semimetals where the band crossings are isolated points 
with four-fold\texttt{\small/}two-fold degeneracy and the quasi-particle excitations disperse
linearly and obey the Dirac\texttt{\small/}Weyl equations. In the the second class, degenerate dispersive lines 
appear either as an open or a closed nodal line across the Brillouin Zone (BZ) and are divided into
type-I, type-II, and hybrid topological NLs~\cite{PhysRevB.98.115164}.
The third class of TSMs involve band crossings that form a nodal surface 
in the BZ~\cite{Liang085427,Zhong7232,Wu115125,Xu205310,Topp041073,Fueaau6459}.
%
%However, in this scope, the material realization is rather scarce, and the band theory has 
%proposed only a few nodal surface semimetals~\cite{Fueaau6459}.

Recent developments in the realization of 2D topological materials will pave the way for
studying exotic quantum phenomena at the nanoscale that might lead to novel quantum devices.
2D NLSMs are new members of this family that have been the focus of attention in recent studies.
Besides the important properties of these systems including non-dispersive Landau energy levels~\cite{rhim2015landau}, 
high-temperature surface superconductivity~\cite{kopnin2011high},
and specific long-range Coulomb interactions~\cite{huh2016long},
one of the most notable characteristics of 2D NLSMs is that their topological features can be revealed by 
the angle-resolved photoemission spectroscopy (ARPES) measurements~\cite{zhou2018coexistence}.
In spite of the large number of discovered 3D TSMs, there are a limited number of 2D materials exhibiting NLSMs
~\cite{jin2017prediction,li2018nonsymmorphic,zhou2018coexistence,zhong2019two,wu2019hourglass}.
Some examples include 2D Lieb lattice~\cite{yang2017dirac,Feng2020experimental}, monolayer borophene~\cite{gupta2018dirac}, 
honeycomb-Kagome lattice~\cite{lu2017two}, 
transition metal chalcogenide monolayers~\cite{jin2017prediction}, Cu$_2$Si~\cite{feng2017experimental}, 
and CuSe~\cite{gao2018epitaxial} monolayers,
where the two last cases are experimentally realized.
Therefore, the exploration of new 2D NLSM materials is of great interest.

The bulk structure of AlB$_2$-type materials are known
to be 3D TSMs and exhibit specific properties including superconductivity in
MgB$_2$~\cite{nagamatsu2001superconductivity} and ZrB$_2$~\cite{gasparov2001electron},
excellent thermoelectricity in AlB$_2$~\cite{sharma2011comparative}
and MgB$_2$~\cite{putti2002electron}, and super-hardness in OsB$_2$~\cite{Cumberland2005Osmium}.
Moreover, novel topological states including triple point, nexus, and nodal links has been verified
experimentally and theoretically in TiB$_2$ and ZrB$_2$
~\cite{Zhang2017Coexistence,Liu2018Experimental,Feng2018Topological,PhysRevB.97.201107,lou2018experimental},
where, the dominant features of the energy bands near the crossing points come from the
Ti- and Zr-3d states and the appeared NLs are protected by mirror reflection symmetries
~\cite{Zhang2017Coexistence,Feng2018Topological}.
Alternatively, a topological Dirac nodal line (DNL), dispersed along the K-H direction
and protected by the combination of inversion and TR symmetries~\cite{jin2019topological}, 
has been observed in the conventional high-temperature superconductor MgB$_2$ and 
its non-superconducting sister AlB$_2$~\cite{jin2019topological,takane2018observation}.
In contrast to TiB$_2$ and ZrB$_2$, the DNLs in AlB$_2$ and MgB$_2$ originate from the B-2p electrons. 
As a result, the spin-orbit coupling (SOC) in these materials is negligible and consequently 
the involved topological nodal lines are experimentally feasible~\cite{takane2018observation}.

It has been shown that the mechanical cleavage of AlB$_2$ flakes may lead to 
highly single-crystalline 2D layers of this material~\cite{humood20182d}. 
In addition, recent ARPES measurements on a typical cleaved (001) surface of AlB$_2$ 
revealed the feasibility of both B- and Al-terminations~\cite{sunko2020surface}.  
In a similar work, a B-terminated monolayer of AlB$_2$ was successfully 
synthesized on Al(111) via molecular beam epitaxy~\cite{geng2020}.
These experimental works, evidence the feasibility of aluminum diboride thin films
with the desired terminations.
Herein, we study three atomic layers AlB$_2$ thin films
with the Al- and B- termenations, leading to the Al$_2$B$_2$ and AlB$_4$ stoichiometires,
by using the first-principles calculations, group 
theory analyses and effective Hamiltonian models (including WTB and continuum models).
%Interestingly, they showed that in this material the in-plane mirror symmetry of the bulk 
%structure is preserved in the Mg-terminated monolayers, either freestanding or on Mg-substrate~\cite{bekaert2017free}.
%Thus, utilizing such a substrate as a base material does not harm
%the mirror symmetry-related phenomena in the material and
%it would be feasible to trace them in experimental works.

After confirmation of the structural stability of these compounds, 
we investigate the topological aspects 
of the characteristic band structures. 
Then, we present a thorough
group theory analysis to explain the protection mechanism behind the
emergence of various  Dirac points and NLs.  Meanwhile, we also provide
effective continuum models using the \textit{method of invariants}~\cite{bir1974symmetry}
to systematically reproduce 
the momentum distribution of DNLs  and responsible
low-energy band dispersions around these Dirac nodes.
Our results indicate that there exist several type-I 2D NLs in a 
relatively small energy range around the Fermi levels of Al$_2$B$_2$ and AlB$_4$
which enclose high symmetry points in the BZ. Very interestingly,
for the Al$_2$B$_2$ configuration  there is an NL that encloses the K point and
crosses the Fermi level with a dispersion of  $\sim$0.6~eV. 
Most strikingly, we realize the coexistence of an open 
type-II 2D nodal line with other Dirac nodal features in AlB$_4$ configuration which 
arises from the intersection of those energy bands that 
contributes to determining the superconducting behavior in this material.
Our predictions not only enrich the family of  2D NLSMs,
but also highlight the potential of these topological materials
as promising candidates to explore   exotic properties in dispersive
2D Dirac nodal lines and the interplay of the superconductivity and topological Dirac nodal line states~\cite{Campi2021prediction}.
\begin{figure}[t!]
    \centering
    \includegraphics[width=.48\textwidth]{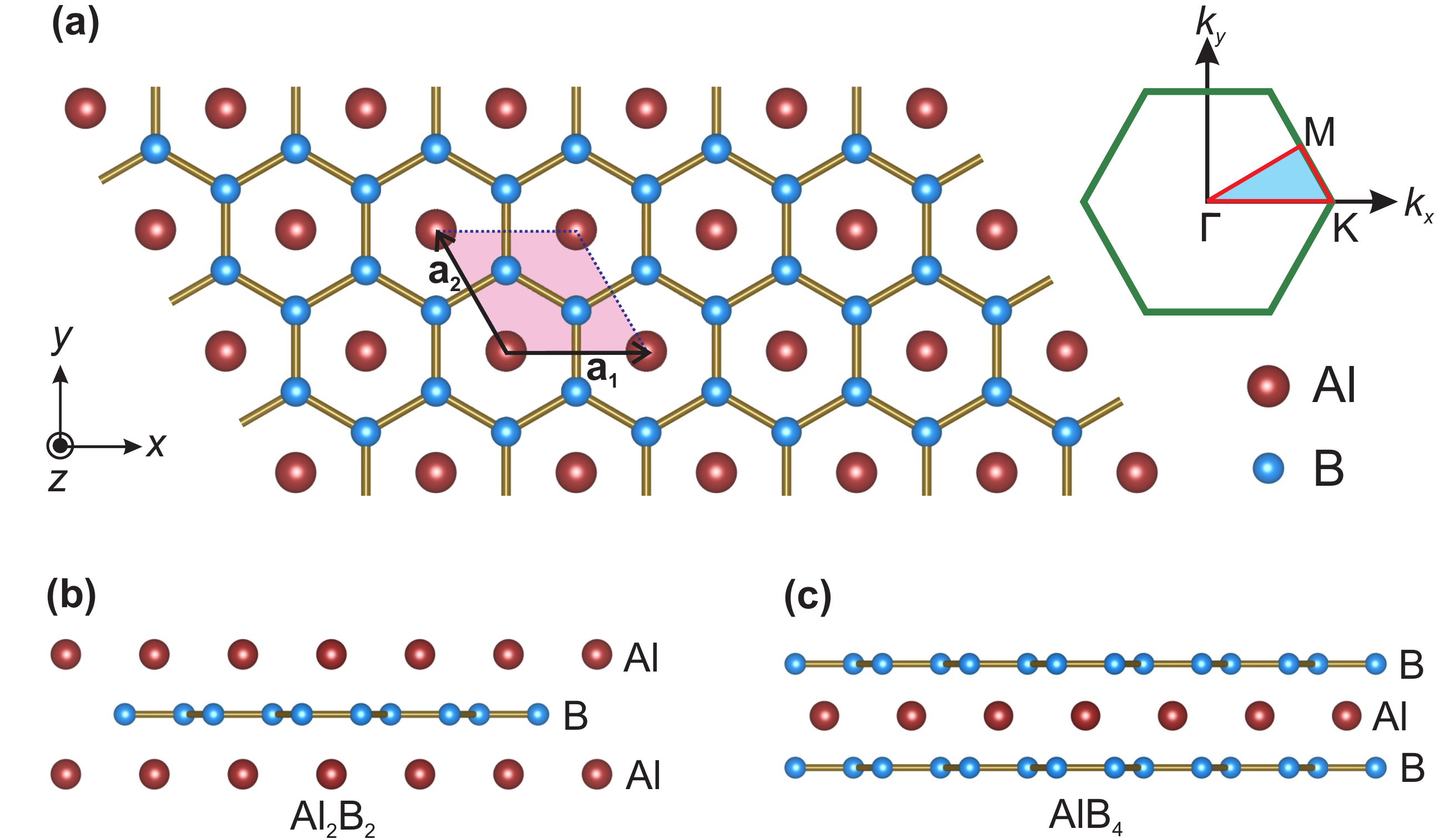} \\
    \centering
    \hspace{-0.7 cm}
    \includegraphics[width=.24\textwidth,valign=t]{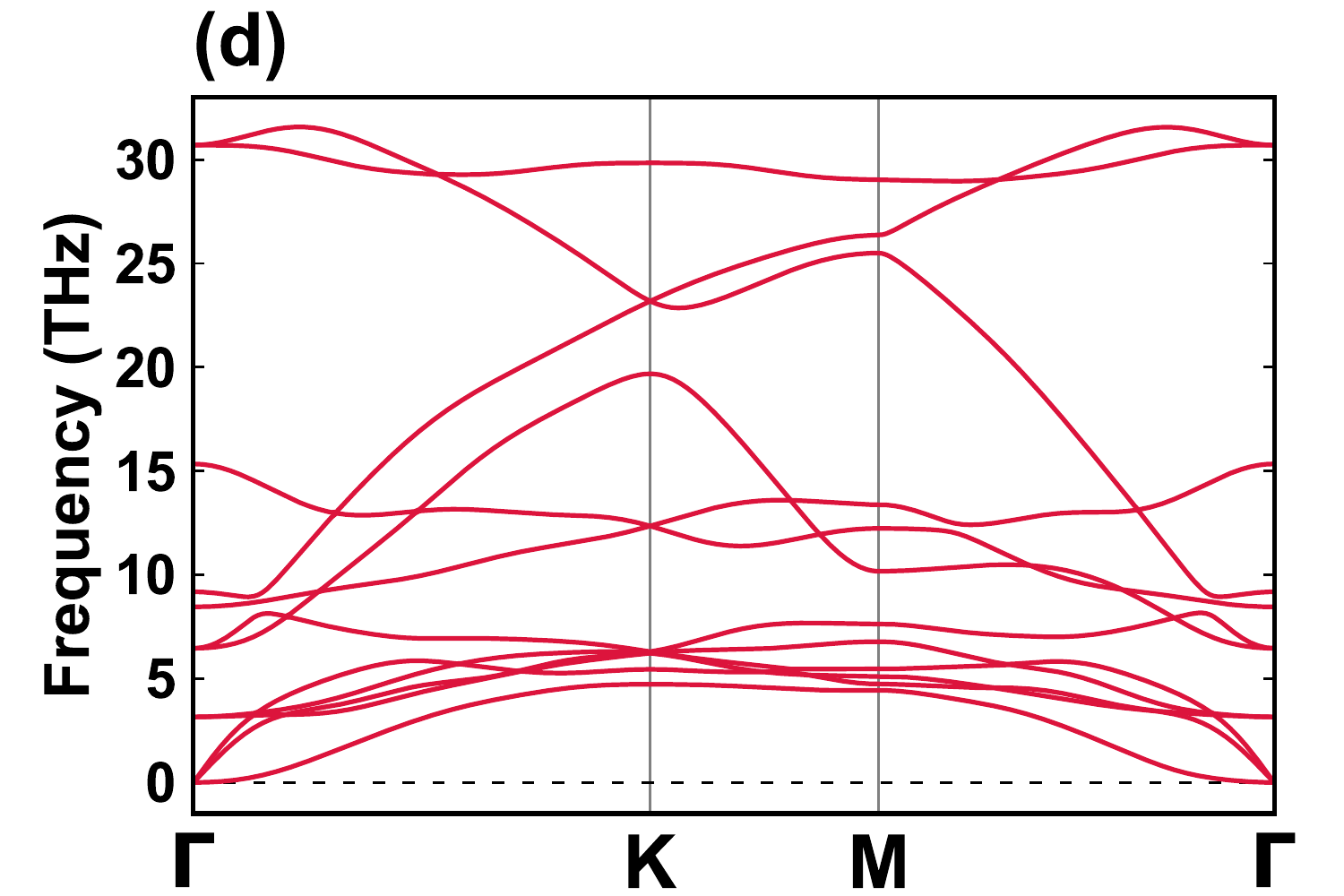} 
    \hspace{-.2 cm}
    \includegraphics[width=.24\textwidth,valign=t]{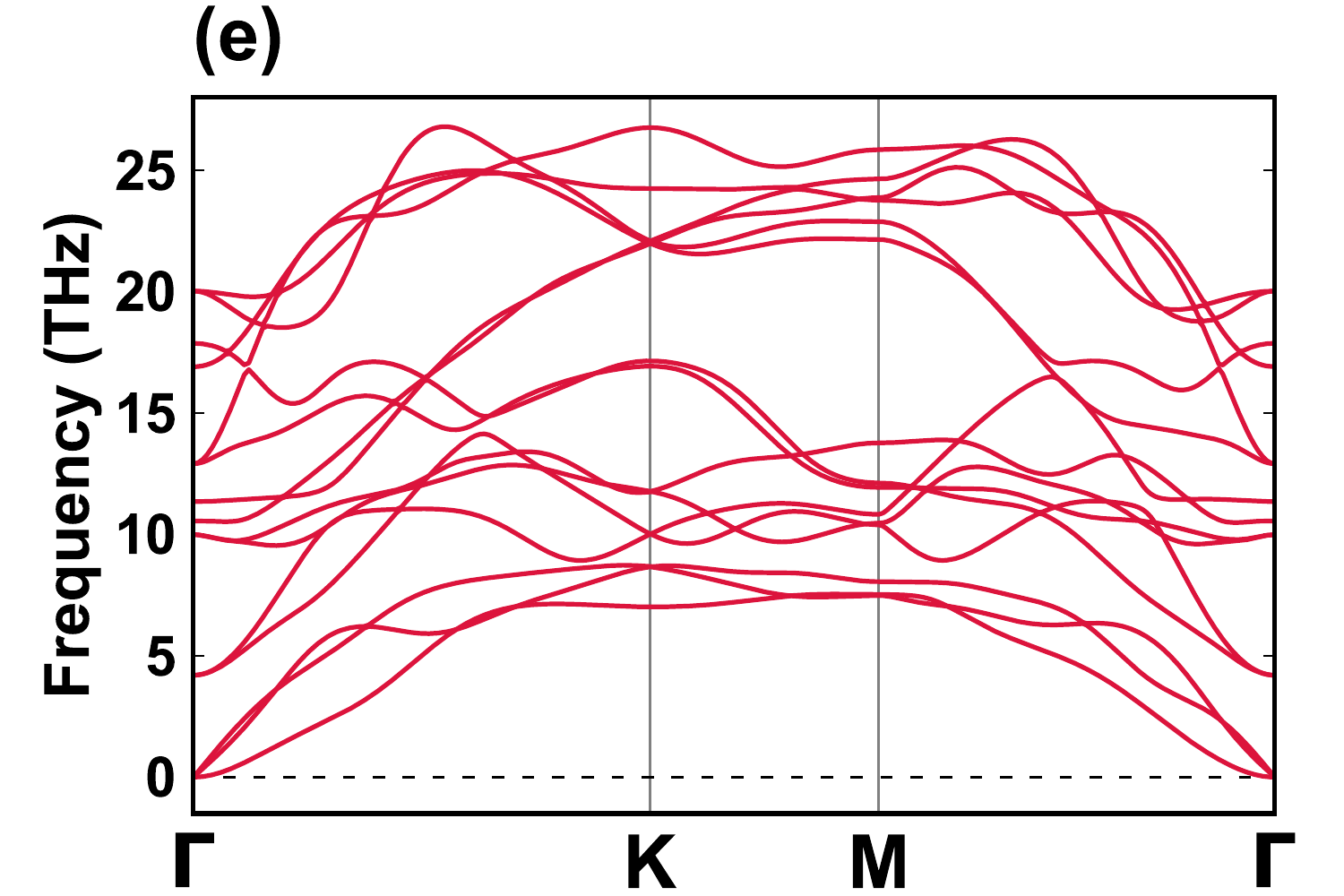} 
    \hspace{-.2 cm}
    \caption{
	     (a) Top view and BZ of aluminum diboride thin films with lattice vectors $\bm a_1$ and $\bm a_2$.
	     (b) and (c) Side views of thinnest films of Al-B$_2$-Al and B$_2$-Al-B$_2$ configurations.
	     (d) and (e) Phonon frequencies of Al$_2$B$_2$ and AlB$_4$.
	    } 
 \label{fig:lattice-phonon} 
\end{figure}
\\
\textbf{\large {Results and discussion}}\\
\textbf{Crystal structures and phonon dispersions.}
Depending on the (001) surface termination,
one may consider two configurations for the thinnest AlB$_2$ films with the in-plane mirror symmetry,
which is crucial for the emergence of nodal line in the system.
Figure~\ref{fig:lattice-phonon}(a) shows the similar top view of these two configurations
where the lattice vectors $\bm a_1$ and $\bm a_2$ indicate their surface unit cells.
The first pattern of the atomic layers lies within the hypo-stoichiometric Al$_{1+x}$B$_2$ films
and consists of a boron layer intercalated between two aluminum layers with 
the Al-B$_2$-Al stacking (Fig.~\ref{fig:lattice-phonon}(b)). 
The second one is the thinnest hyper-stoichiometric AlB$_{2+x}$ film with a sandwich 
B$_2$-Al-B$_2$ configuration (Fig.~\ref{fig:lattice-phonon}(c)), 
where the two boron layers form an AA-stacking order enclosing a triangular aluminum layer.  
%The symmetry group of these 2D ultrathin films is one of the eighty layer groups.
The crystal structure of our hypo- and hyper-stoichiometric thin films,
refered to as Al$_2$B$_2$ and AlB$_4$, respectively,
belongs to the layer group $P6\texttt{\small/}mmm$ (No.80) with the corresponding point group $D_{6h}$~\cite{litvin2012character}.
%The thinnest AlB$_2$ film consists of a honeycomb boron plane and a triangular
%plane of Al atoms. This configuration adopts the layer group $P6mm$ (No.77), and the corresponding 
%point group is $C_{6v}$~\cite{litvin2012character}.
Our structural optimizations lead to the in-plane lattice constants 2.994~{\AA} and 2.949~{\AA}
for the AlB$_4$  and Al$_2$B$_2$ films, respectively.  
To evaluate the dynamic stability of these structures, we calculated
and considered their phonon spectra.
The obtained phonon band structures, presented in Figs.\ref{fig:lattice-phonon}(d) and (e), 
clearly indicate absence of any imaginary phonon mode, thus implying their dynamical stability.\\
Having established the stability of these configurations, we now turn our attention to 
the electronic and topological properties.
As both the AlB$_4$ and Al$_2$B$_2$ structures are composed of light elements, 
the SOC is quite weak in these materials and one would
expect to observe a definite sign of possible Dirac features in experimental measurements.
Therefore, we shall ignore the spin of electrons throughout our calculations.\\
\textbf{DNLs in Al$_2$B$_2$.}
We begin by investigating the electronic and topological properties 
of Al$_2$B$_2$ configuration. As mentioned before, the point group of this  structure is $D_{6h}$ 
which contains 24 symmetry elements.
We have shown in Fig.~\ref{fig:mirror-planes}(a) those symmetry elements
that are important for our purpose.
They are the two different sets of twofold symmetry axes  ($3C'_2$ and $3C''_2$), 
two different sets of vertical mirror planes ($3M_\sigma$ and $3M_\tau)$, and one horizontal  symmetry plane ($M_h$).
Figure~\ref{fig:Al2B2-band-structure}(a) depicts the  electronic  band structure  of
%atomic orbital-projected band structure
%and the corresponding  \hl{density of states}  
Al$_2$B$_2$  along the highly symmetric
directions (M-K-$\Gamma$-M) in the  BZ.
\begin{figure}[t!]
    \centering
    \hspace{-0.7 cm}
    \includegraphics[width=.45\textwidth,valign=t]{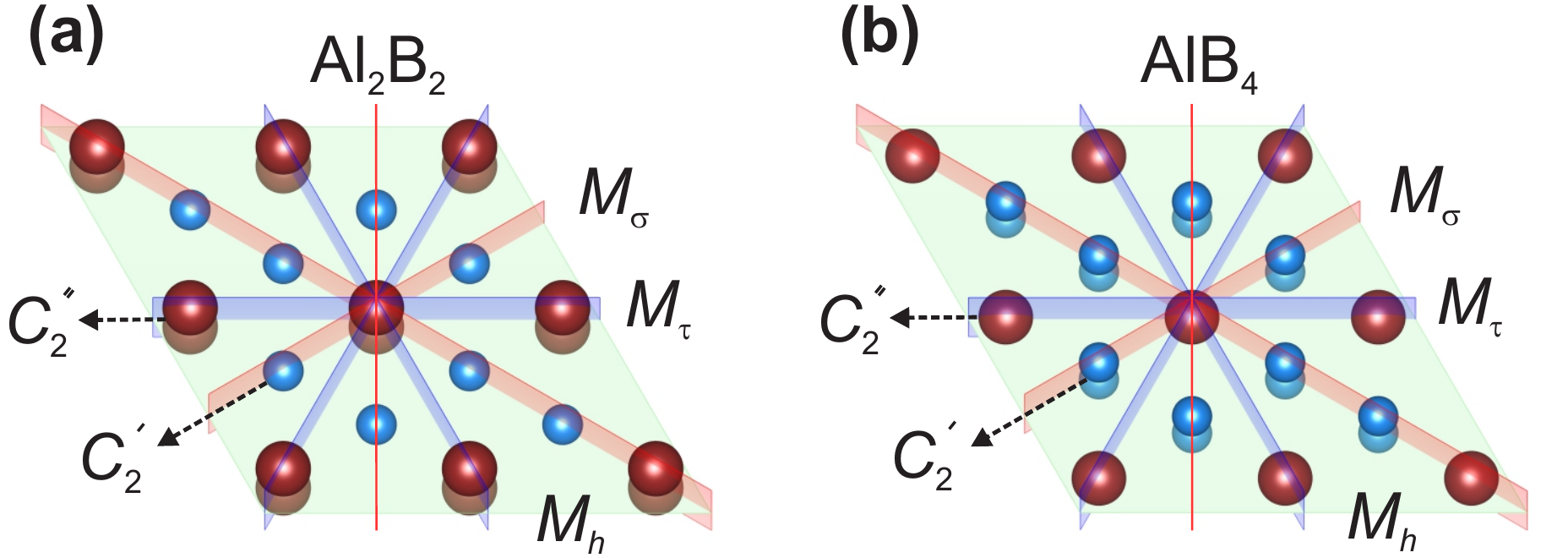}
    \caption{
             Three types of mirror planes $M_\sigma$, $M_\tau$, and $M_h$ for  (a) Al$_2$B$_2$ and (b) AlB$_4$
             configurations, as indicated by the red, blue, and green planes, respectively. Two of rotational axes $C'_2$ and $C''_2$  are shown by dashed arrows.
	    } 
 \label{fig:mirror-planes} 
\end{figure}
As seen
from Figs.~\ref{fig:Al2B2-band-structure}(b) and (c),  the atom-projected
and orbital-projected bands show that the metallicity of 
Al$_2$B$_2$ comes from the hybridization among $s$ and $p$ electron
valence shells of both Al and B atoms.
Interestingly, in the energy range from -2.4 to 1.2 eV,  we find multiple band crossing
features (consisting of Dirac points (DPs) and nodal points (NPs))
that offer a fruitful line of investigation for 2D topological NLs in this configuration.
A close inspection of the band structure reveals that in this 
energy window there exist two Dirac points and 
five type-I  2D NLs. We mark the two Dirac points as DP1 and DP2 and indicate
the nodal points as NP$i$ and NP$i'$ ($i$ runs from 1 to 5) 
that are the pertinent points in the nodal loop NL$i$ (see Fig.~\ref{fig:Al2B2-band-structure}(a)).
The formation of multiple topological nodal loops and Dirac fermions in this 
band structure suggests that the understanding of the mechanism behind it has to be questioned.
Therefore, what follows is a discussion to demystify the
reason for the observation of the  introduced 2D NLs  and Dirac 
points utilizing the group theory analysis, TB approximation, and method of invariants.
\begin{figure}[t!]
    \centering
    \hspace{-0.5 cm}
    \includegraphics[width=.48\textwidth,valign=t]{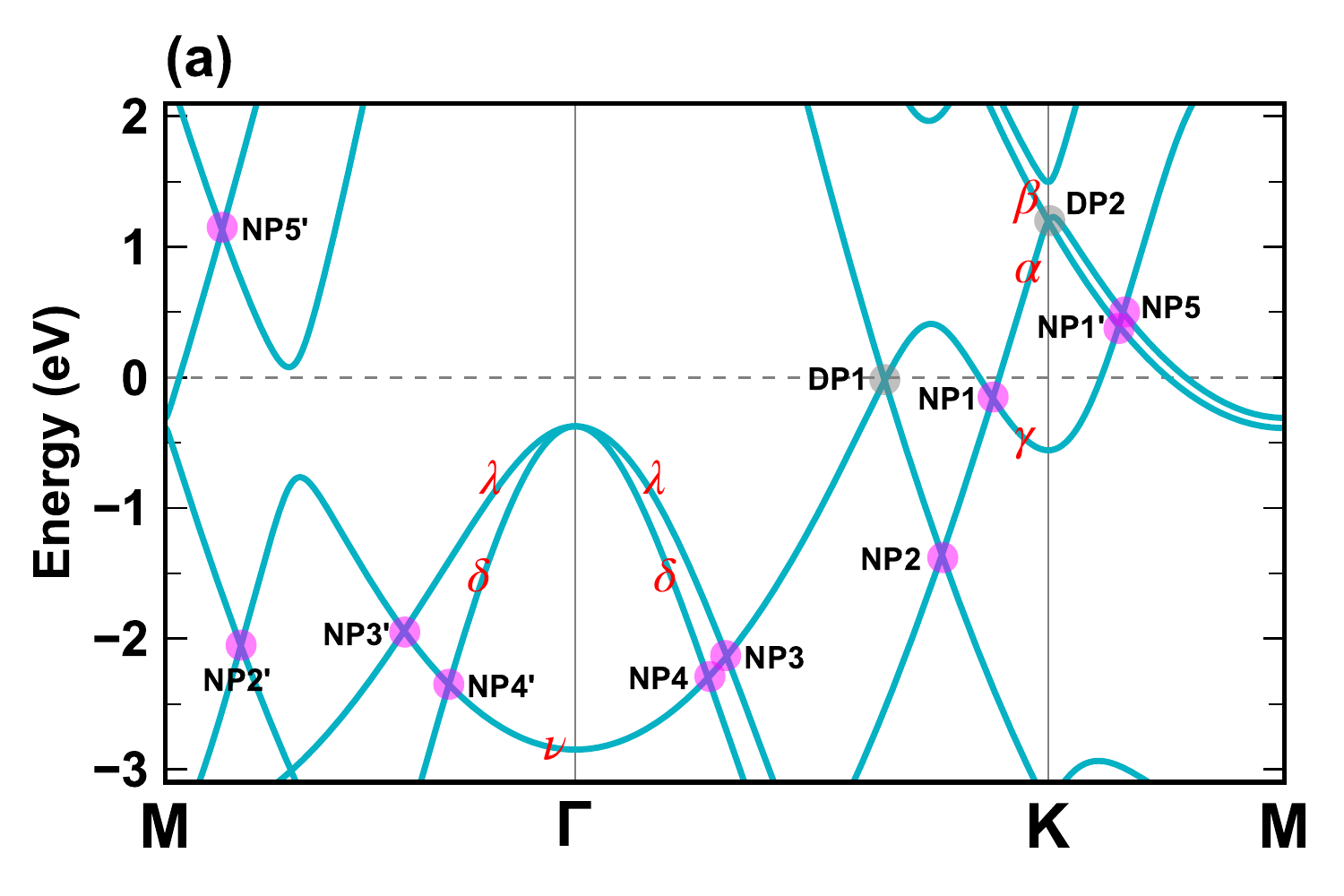}  \\ 
    \hspace{0.1 cm}
    \includegraphics[width=.235\textwidth,valign=t]{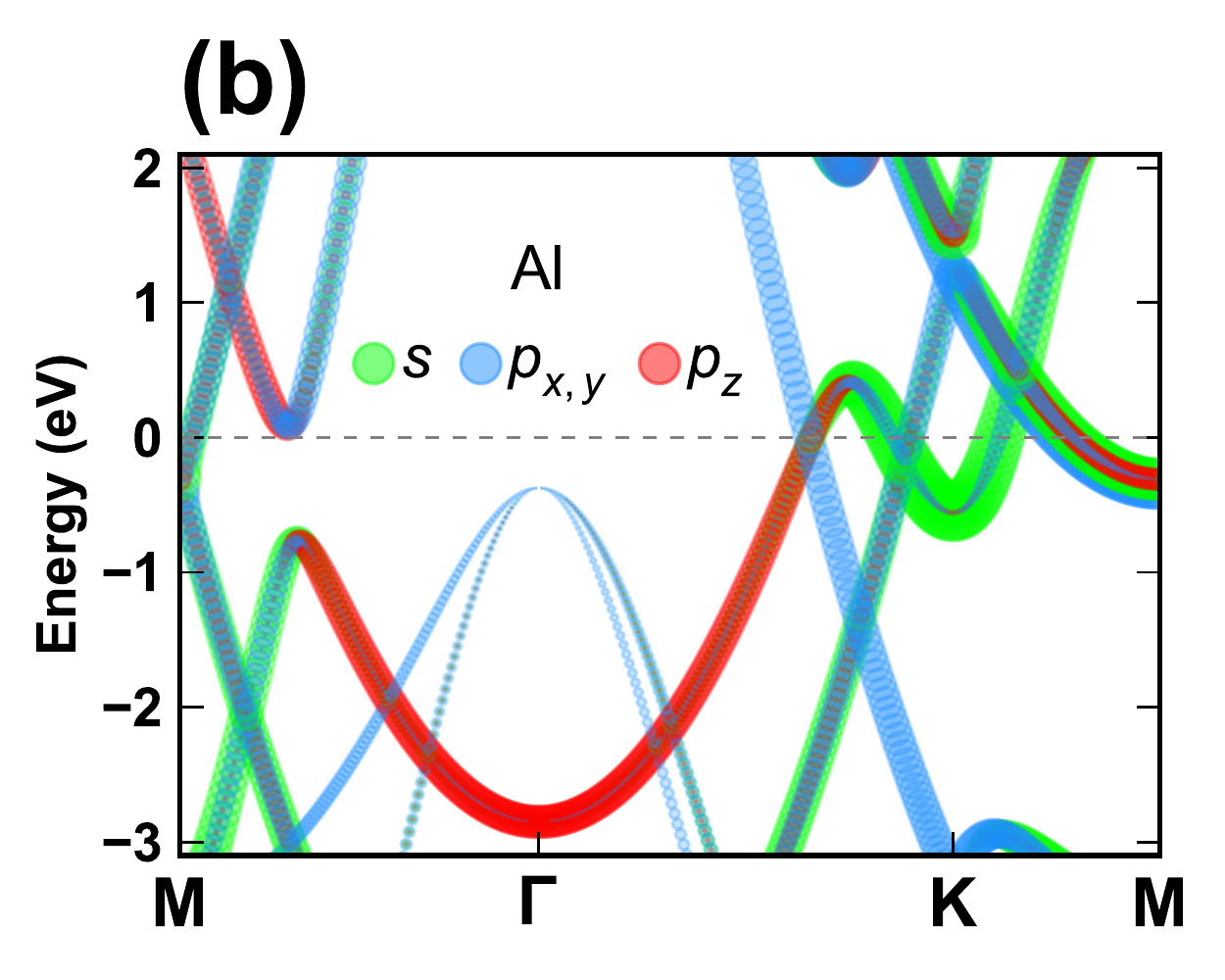}
    \hspace{-0.2 cm}
    \includegraphics[width=.235\textwidth,valign=t]{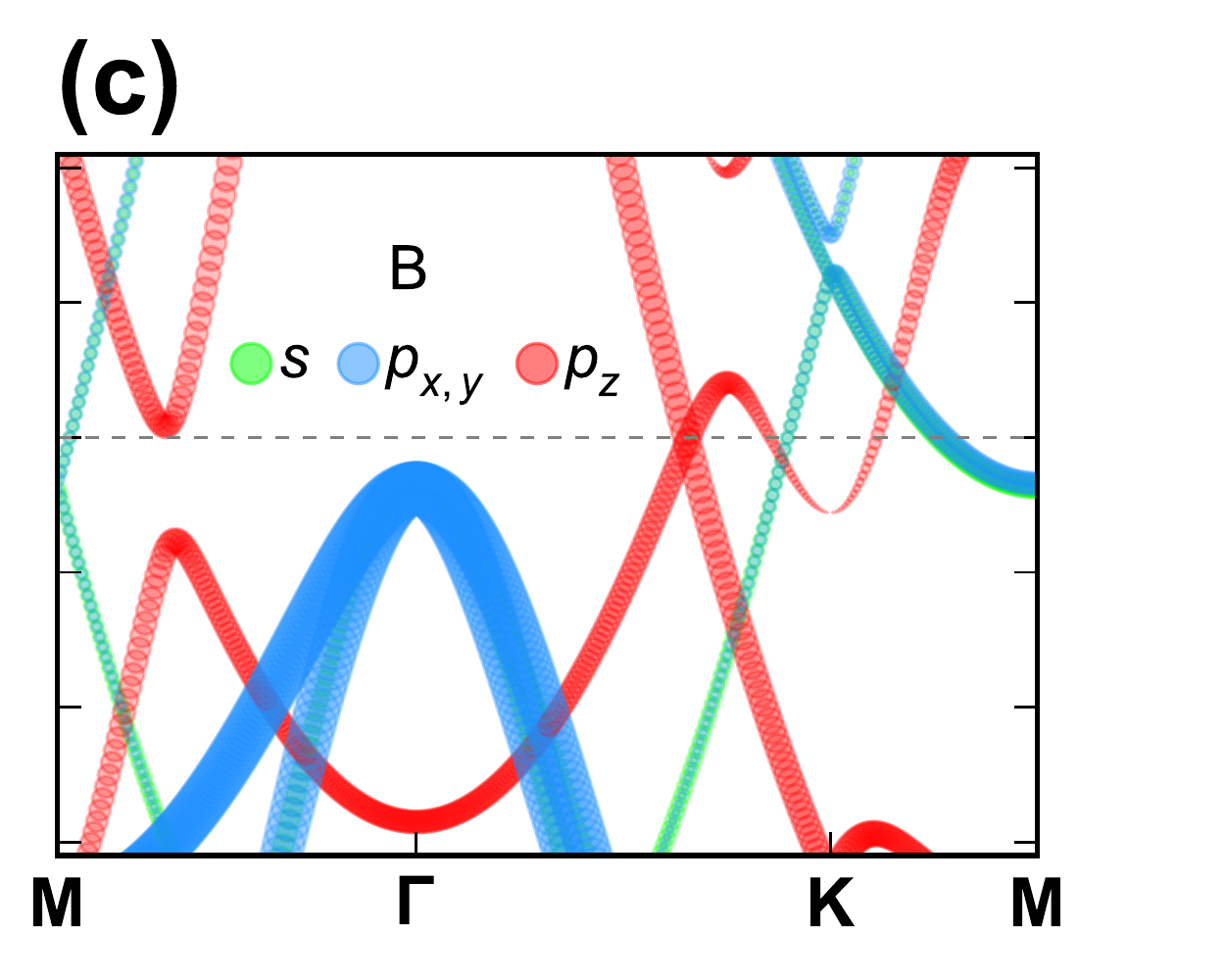}
    \caption{
             (a)  The  calculated  electronic  band structure  of Al$_2$B$_2$ configuration.  Energy  dispersion 
             display multiple band crossing features including both Dirac and nodal points as presented by magenta circles.
             Special bands are marked by Greek letters.
             (b) and (c) Atom-projected band structures for Al$_2$B$_2$ contributed by Al and B atoms, respectively.  
             The size of dots is proportional to the weight of contributed orbitals.
            } 
 \label{fig:Al2B2-band-structure} 
\end{figure}
%$k\cdot p$ method, 
\iffalse us begin our discussion by plotting 3D energy bands of
monolayer  Al$_2$B$_2$ which provides a better insight in the
whole BZ. To do this,  we construct a TB Hamiltonian in the basis of maximally localized  Wannier functions (?MLWFs).
Next,  we diagonalize it to provide the 3D energy bands whose 
intersections form the mentioned DNLs and Dirac points as shown in Figs..\fi
First of all, to gain a better insight into the band structure of monolayer  Al$_2$B$_2$ in the
whole BZ, we have constructed a TB Hamiltonian in the basis of maximally localized  Wannier functions.
The comparison of the band structures of  Al$_2$B$_2$ as
calculated by density functional theory (DFT) and the WTB Hamiltonian is shown in Fig.~\ref{fig:Al2B2_dft-wann-comparison}. One can see that in the 
mentioned energy range the Wannier band structure exactly matches the DFT results.
Next, we have diagonalized our Hamiltonian in the whole BZ to provide the 3D energy bands whose 
intersections form the mentioned 2D DNLs. Figures \ref{fig:Al2B2-3d-plots}(a)-(d) depict zoomed-in band structures in the regions 
of band crossing near nodal points NP$i$ and NP$i'$ (see top views in Figs.~\ref{fig:Al2B2-3d-plots}(e)-(h)).
From the transparent top views of bands surfaces,
it is clear that the two points NP$i$ and NP$i'$  are residing on a 2D nodal loop NL$i$.
The extensions of nodal points are displayed 
throughout 3D energy-momentum spaces which vividly describe the form of characteristic topology of each NL.
As seen, NL1 and NL5 surround the K and M points, respectively and
NL2, NL3, and NL4 are three concentric nodal loops centered around the $\Gamma$ point.
\iffalse
Also, we have presented in Figs.... the color maps of the local gap between crossing bands which give explicitly
the 2D momentum distributions of the nodal loops and the location of Dirac points.
\fi
These Dirac nodal loops have dispersion at different energy levels in the BZ such that
cover a large  portion of the mentioned energy window.\\
Let us now turn to the mechanism by which the Dirac points DP1 and DP2 emerge.
As shown in Fig.~\ref{fig:Al2B2-band-structure}(a), the DP1 point is located along the $\Gamma$-K direction and 
25~meV below the Fermi level.  The little group for Al$_2$B$_2$ configuration along
this high symmetry direction is $C_{2v}$ with two perpendicular mirror reflections 
planes; $M_{\tau}$ and $M_{h}$ which their intersection introduces the twofold rotation axis $C_{2}$.  
Our symmetry analysis shows that the irreducible representations (IRs) of the crossing bands around the
DP1 point are $\Gamma_3$ and $\Gamma_4$ (Fig.~\ref{fig:Al2B2_dft-wann-comparison}).
Since the IRs $\Gamma_3$ and $\Gamma_4$ under the $M_{\tau}$ or $C_{2}$  symmetry 
operation have opposite parities, the 
two bands around the DP1 point do not interact. As a result, six symmetry
protected Dirac points DP1 emerge in the whole BZ due to the star of $k$.  
The  Dirac point DP2 with an n-type character is located at the K point which stems
from the underlying point group $D_{6h}$  
of the lattice in a similar manner to the formation of Dirac cones in graphene. \\
We now begin to consider the mechanism which generates NL1.
An important feature of  this nodal line is  that it crosses the Fermi level  
and disperses from about $-0.2$ to $+0.4$~eV (see Fig.~\ref{fig:Al2B2-band-structure}(a)).
For more convenience,  in the vicinity of the nodal points NP1 and NP1$'$
we label the responsible electronic bands as $\alpha$, $\beta$, and $\gamma$.
\begin{figure}[t]
    \centering
    %\hspace{0.1 cm}
    \includegraphics[width=.48\textwidth,valign=t]{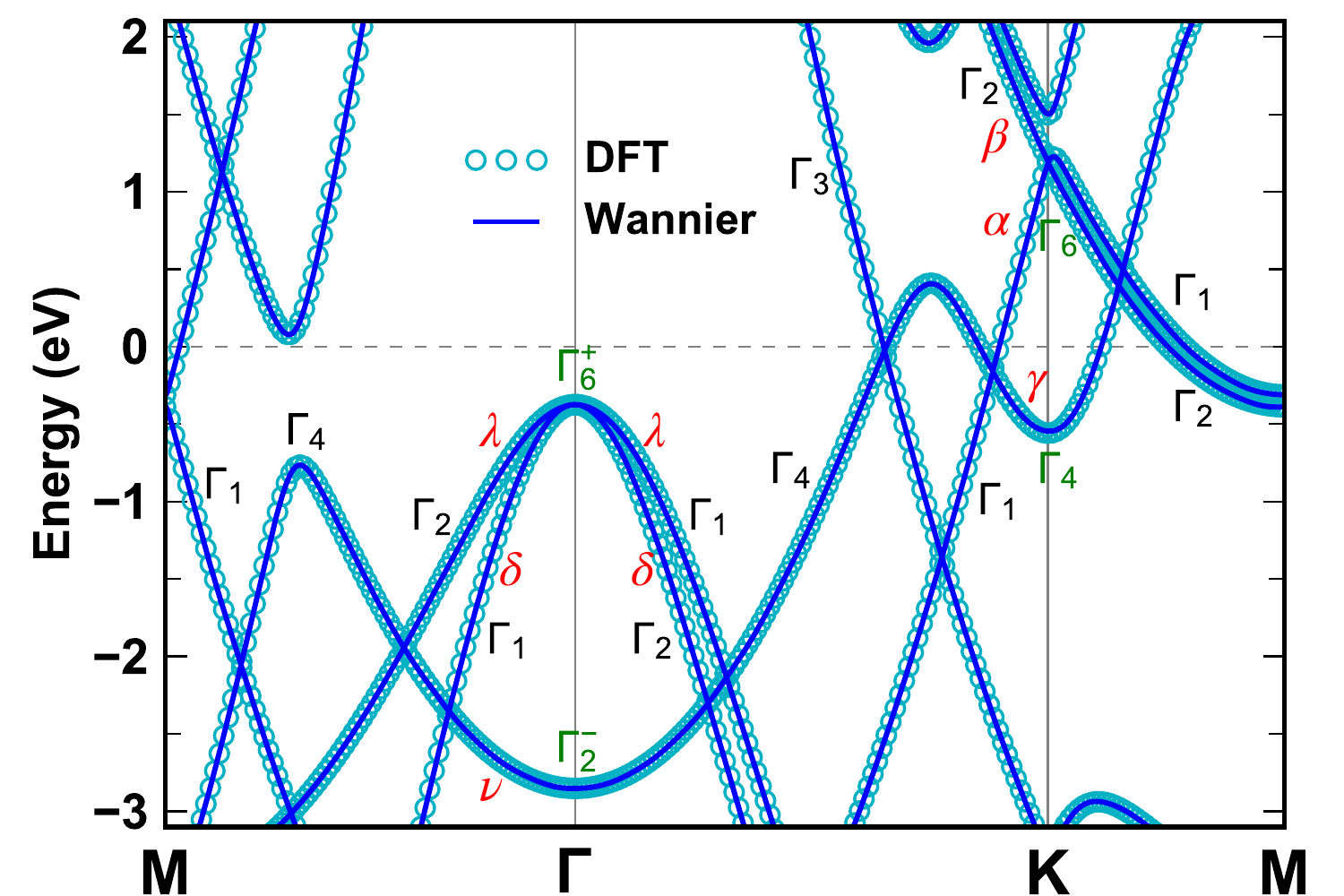} \\ %\quad
    %\hspace{-0.2 cm}
    %\includegraphics[width=.4\textwidth,valign=t]{img/AlB4_DFT_WANN_compare_bands.pdf} %\quad 
    \caption{ Comparison of the band structures of Al$_2$B$_2$ along highly symmetric directions as calculated by DFT and the WTB Hamiltonian. 
    The Wannier band structure exactly matches the DFT results. The IRs of energy bands along these symmetry
    paths and at high symmetry points K and $\Gamma$ are also shown.
            } 
 \label{fig:Al2B2_dft-wann-comparison} 
\end{figure}
\begin{figure*}[t]
    \centering
    \includegraphics[width=1.0\textwidth,valign=t]{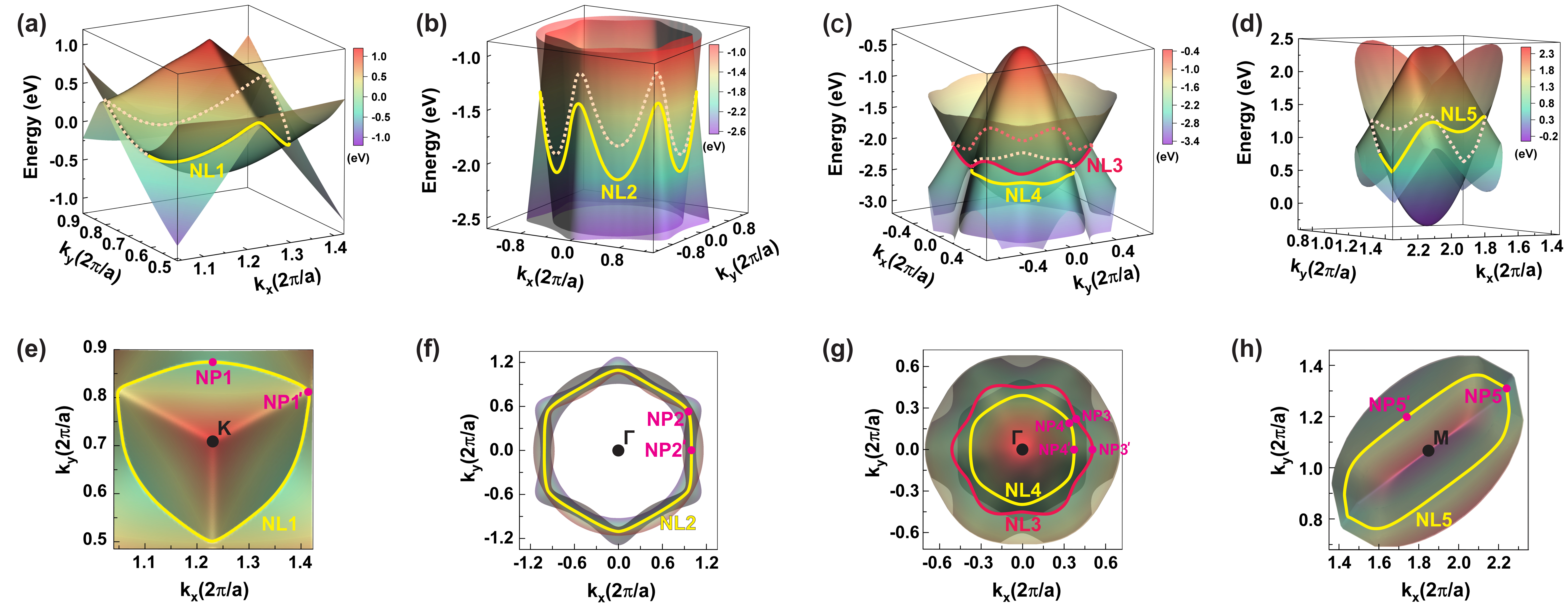} %\quad
    \caption{
             (a)-(d) 3D visualizations of energy bands obtained from  Al$_2$B$_2$   
             WTB Hamiltonian. Bands whose crossings lead to the formation of nodal lines around
             the K, $\Gamma$, and M points are shown. The extensions of nodal points can be seen vividly
             throughout 3D energy-momentum spaces which form the characteristic topology of each NL.
             Color bars represent the energy of electrons.
             (e)-(h) The corresponding top views of (a)-(d). The momentum distributions of gapless nodal points are shown. 
             Note that in Figs.(b) and (f) some data values were removed for better visualization.
            } 
 \label{fig:Al2B2-3d-plots} 
\end{figure*}
\begin{figure}[t]
    \centering
    \includegraphics[width=.5\textwidth,valign=t]{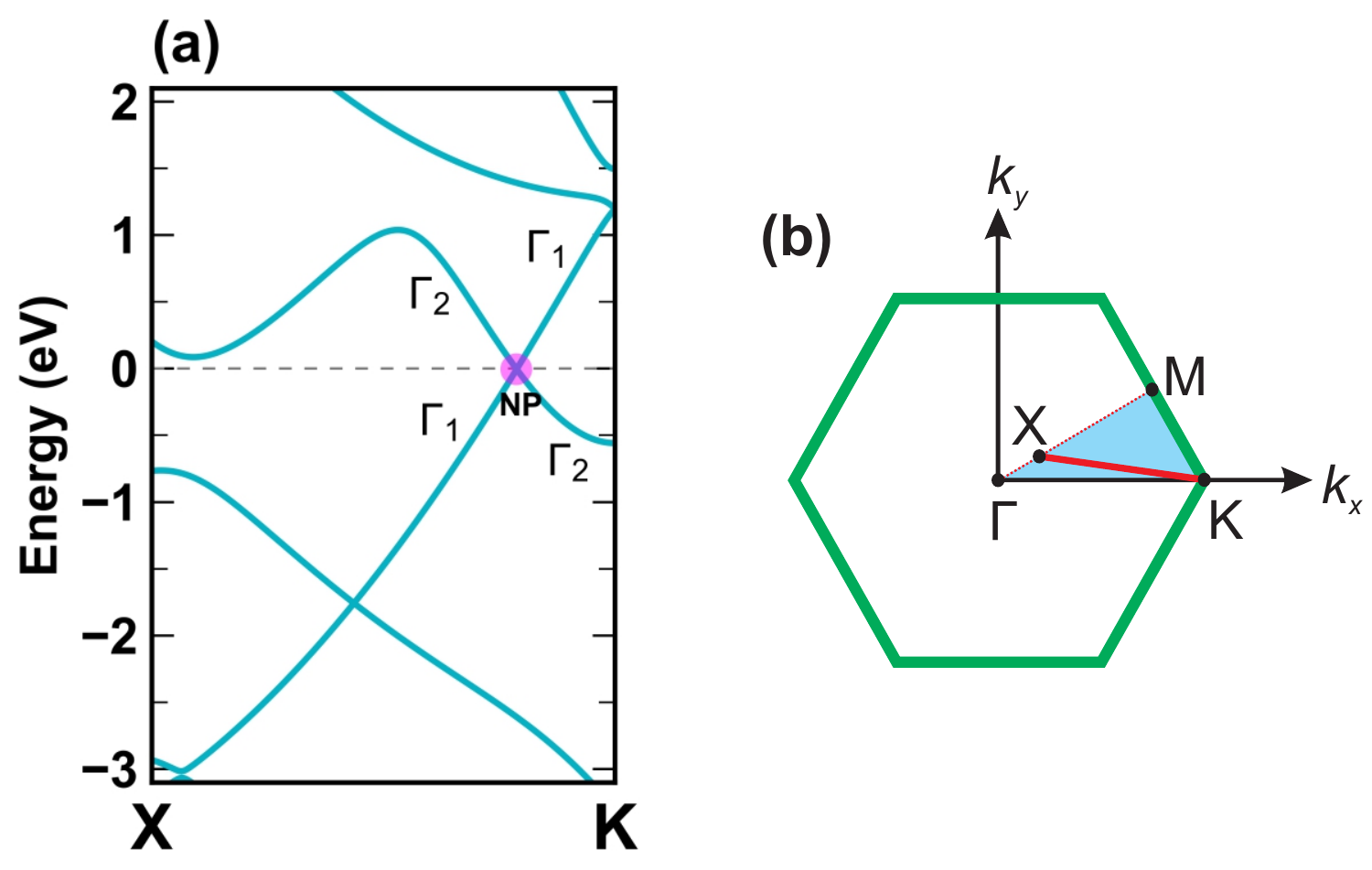} 
    \caption{
             (a) Electronic band structure of Al$_2$B$_2$ along the typical
             low-symmetry path X-$\Gamma$ in the BZ as shown in (b).
            } 
 \label{fig:Al2B2_BS_XK} 
\end{figure}
%Near the K point, the three bands whose crossing results into this nodal line
%are denoted by $\delta$, $\lambda$, and $\nu$.
%As seen in Fig., near the Fermi level the electronic states of the bands which are
%responsible for the nodal points NP1 and NP1$'$  are mainly decided by Al-s and Al-p$_{x,y}$.
As seen from Figs.~\ref{fig:Al2B2-band-structure}(b) and (c), the electronic states of these bands near the Fermi level in the vicinity 
of the nodal points NP1 and NP1$'$  are mainly decided by Al-s and Al-p$_{x,y}$.
Similar to the $\Gamma$-K path, the little group along the K-M direction is also $C_{2v}$
with the mentioned symmetry operations. Group theory analysis shows that the $\alpha$, $\beta$, and $\gamma$ bands
belong to three different $\Gamma_1$, $\Gamma_2$ and $\Gamma_4$ IRs, respectively (Fig.~\ref{fig:Al2B2_dft-wann-comparison}).
\iffalse
Similar to the case of AlB$_4$, the little group for 
Al$_2$B$_2$ configuration along the $\Gamma$-K and K-M directions is also $C_{2v}$
with two perpendicular mirror reflections planes; $M_{\sigma}$ and $M_{h}$ which their intersection
introduces the twofold rotation axis $C_2$. Group theory analysis show the $\delta$, $\lambda$, and $\nu$ bands
belong to three different $A_2$, $A_1$ and $B_2$ IRs, respectively.
\fi
Along the $\Gamma$-K direction, the  $\alpha$ and $\gamma$ bands 
have opposite eigenvalues either with respect to $C_2$ or $M_h$  symmetry operations. These opposite parities ensure the decoupling of 
the $\alpha$ and $\gamma$ bands leading to the formation of topologically protected NP1. 
On the other hand,  along  the  symmetry line K-M, the 
opposite parities of $\beta$ and $\gamma$  bands under either $M_\tau$ or $M_h$ operations allow the emergence of topological point 
NP1$'$. Besides the justification of the two nodal points NP1 and NP1$'$
along high-symmetry paths, we also examine the mechanism of the other touching points
in the NL1.  Figure~\ref{fig:Al2B2_BS_XK} illustrates the band structure of Al$_2$B$_2$ along a typical low-symmetry
line X-K where X is a point between the $\Gamma$ and M points.
Due to the  mirror reflection symmetry $M_h$
of monolayer Al$_2$B$_2$,  we found that the little group  of the X-K  line
is $C_s$ and the two crossing bands near the Fermi level belong to different
IRs $\Gamma_1$ and $\Gamma_2$ as marked in Fig.~\ref{fig:Al2B2_BS_XK}(a). This is  valid  for the set of the mentioned
typical paths. These different representations
have opposite parities with respect to $M_h$,  giving rise to a distribution of touching points
that construct the 2D topological NL1 (see Figs.~\ref{fig:Al2B2-3d-plots}(a) and (e)).\\
To further analyze the structure of NL1 and the corresponding crossing bands, 
we use group theory to derive a matrix Hamiltonian with invariant
expansion method~\cite{bir1974symmetry} which describes the electron states in the vicinity of the K point.
This method enables us to systematically incorporate the 
allowed terms up to any desired order of wave vector $\bm k$.
The little group of the wave vector K in Al$_2$B$_2$ is
$D_{3h}$. At this point the two degenerate Bloch states of the 
$\alpha$ and $\beta$ bands transform according to the 2D IR $\Gamma_6$ as denoted by $\psi^{\alpha}$ and $\psi^{\beta}$, while the 
corresponding wave function of the $\gamma$ band ($\psi^{\gamma}$) transforms as IR $\Gamma_4$ . These basis functions allow us to
construct the $3\times3$ matrix Hamiltonian $\mathcal{H}^\mathrm{K}$ around the K point which falls into four blocks
\begin{equation}
  \label{eq:Ham3*3}
  \mathcal{H}^\mathrm{K} = \left( \begin{array}{cc}
   \mathcal{H}_{11} & \mathcal{H}_{12} \\
   \mathcal{H}_{21} & \mathcal{H}_{22}
    \end{array}\right) \, ,
\end{equation}
where each block $\mathcal{H}_{ij}$ has a dimension of $n_i\times n_j$ and $n_i$ ($n_j$) denotes the dimension 
of representation $\Gamma_i$ ($\Gamma_j$).
The diagonal blocks $\mathcal{H}_{11}$ ($2\times2$) and $\mathcal{H}_{22}$ ($1\times1$) describe energy bands that transform
according to the IRs $\Gamma_6$ and $\Gamma_4$, respectively. Each block $\mathcal{H}_{ij}$ depends on a 
tensor operator $\vekc{K}$ which represents a function of the components of the  wave vector $\bm k$.
Bir and Pikus~\cite{bir1974symmetry} showed that for a symmetry element $g$ with matrix representation $ \mathcal{D}(g)$,  
the invariance condition
\begin{equation}
  \label{eq:spatial-invar}
  \mathcal{D}(g) \, \mathcal{H} (g^{-1} \vekc{K}) \, \mathcal{D}^{-1}(g)
  = \mathcal{H} (\vekc{K}),
\end{equation}
leads to
the general form $\mathcal{H}_{ij}$ as
\begin{equation}
  \label{eq:invar}
  \mathcal{H}_{ij} (\vekc{K}) 
  = \sum_{\kappa, \, \nu}
  \koeff{a}{ij}{\kappa\nu}
  \sum_{l=1}^{d_\kappa} {X}_l^{(\kappa,ij)}
  \mathscr{K}_l^{(\kappa,\nu) \, \ast} \, .
\end{equation}
The parameters $\koeff{a}{ij}{\kappa\nu}$ are material-dependent constants which can be determined via a fitting procedure.
$X_l^{(\kappa,ij)}$ are those linearly independent $n_i\times n_j$-dimensional  basis matrices that transform 
as those irreducible representations $\Gamma_{\kappa}$ which occurs in the product $\Gamma_i \times \Gamma_j^\ast$. Here, $l$ denotes
the $l$th basis matrix and runs up to $d_{\kappa}$, the dimension of representations $\Gamma_{\kappa}$. Also, 
$\mathscr{K}_l^{(\kappa,\nu)}$ are the $l$th component of the $\nu$th-order of the tensor operator $\vekc{K}$ which 
transform in a similar way. To construct the basis matrices $X_l^{(\kappa,ij)}$ and an arbitrary order of the tensor 
operators $\mathscr{K}_l^{(\kappa,\nu)}$, one can utilize Clebsch-Gordan coefficients
which have been tabulated in Ref.~\cite{koster1963properties}. The construction procedure of these 
matrices- and tensor-components has been described in Refs.~\cite{geissler2013group,winkler2003spin}.\\
Besides the invariance condition (\ref{eq:spatial-invar}), the TR symmetry of the system also imposes an additional constraint 
on the diagonal blocks $\mathcal{H}_{ii}$~\cite{winkler2003spin}. Therefore,  when using the invariant method to construct 
the effective Hamiltonians, we will additionally consider this constraint to list the
allowed tensor-components $\mathscr{K}_l^{(\kappa,\nu)}$.\\
Let us now proceed by generating the clear form of 
the Hamiltonian $\mathcal{H}^\mathrm{K}$ up to second order in $\bm k$ components
using the introduced method.
As mentioned,  around  the K  point the Bloch functions $\psi^{\alpha}$ and $\psi^{\beta}$
transform according to the IR  $\Gamma_6$  of $D_{3h}$. Also,
the Bloch function $\psi^{\gamma}$ transforms as representation $\Gamma_4$.
Therefore,  from a group theoretical point of view, one may
adopt $\{ x-iy,x+iy \}$ and $\{z\}$ as basis functions to obtain the basis matrices $X_l^{(\kappa,ij)}$.
Choosing this set of basis functions, we obtained the basis matrices as listed
in Table \ref{tab:Al2B2-K} (see \ref{SN1}). 
Likewise, we can obtain the irreducible tensor  components
up to $k_ik_j$ terms that transform accordingly and are allowed by  TR invariance.
According to Herring's rule, considering the TR symmetry
may lead to three different possibilities '$a$', '$b$', and '$c$' for a spinless system~\cite{herring1937effect}.
In case '$a$', no additional degeneracies occur by applying the TR operator. 
However, when the two inequivalent wave vectors   $\bm k$ and $-\bm k$ 
are related via a symmetry element $R$ (known as case '$a_2$'), one
needs to consider the additional condition~\cite{bir1974symmetry} 
\begin{equation}
  \label{eq:timea2}
  \mathcal{T}^{-1} \mathcal{H} (R^{-1} \vekc{K}) \mathcal{T}
  = \mathcal{H}^\ast (f \vekc{K}) = \mathcal{H}^t (f \vekc{K}),
\end{equation}
where $f=\pm 1$ is the parity of $\vekc{K}$ under the TR symmetry.
\iffalse According to the Herring rule, the electron states at the K  and K' points in alb2 belong to the case 'a2'.  Therefore, 
there exists a symmetry element R in which maps these two 
inequivalent points onto each other, and one can consider the additional 
constraint \fi
In Al$_2$B$_2$ the symmetry element $M_\sigma$ maps 
the two inequivalent points K  and K$'$ onto each other. Choosing the
operator $M_\sigma$, the TR operator have the form of $\mathcal{T}=K$~\cite{winkler2010invariant}, where
$K$ is the complex conjugate operator. Therefore the condition (\ref{eq:timea2})
for the K point is written as 
\begin{equation}
  \label{timea2}
   \mathcal{H}^\mathrm{K}(R^{-1} \vekc{K}) = \mathcal{H}^\ast (f \vekc{K}).
\end{equation}
We have listed in Table \ref{tab:Al2B2-K} those tensor components that 
satisfy the condition~(\ref{timea2}). Then,
inserting the tabulated symmetrized matrices and 
tensor components in Eq.~(\ref{eq:invar})  we arrive at
\iffalse
\begin{flalign}
\label{H1122}
  & \mathcal{H}_{11} = [E_0+A(k_x^2 + k_y^2)]\openone+B(k_x\sigma_z-k_y\sigma_x)\notag\\
  & ~~~~~~~~+C[(k_y^2- k_x^2)\sigma_z-2k_xk_y\sigma_x],\\
  & \mathcal{H}_{22}=E'_0+A'(k_x^2 + k_y^2) , \\
  & \mathcal{H}_{12}=\mathcal{H}_{21}=0.
\end{flalign}
\begin{flalign}
\label{H1122}
  & \mathcal{H}_{11} = [E_0+A(k_x^2 + k_y^2)]\openone+B(k_x\sigma_x + k_y\sigma_y)\notag\\
  & ~~~~~~~~+C[(k_y^2- k_x^2)\sigma_x + 2k_xk_y\sigma_y],\\
  & \mathcal{H}_{22}=E'_0+A'(k_x^2 + k_y^2) , \\
  & \mathcal{H}_{12}=\mathcal{H}_{21}=0.
\end{flalign}
\fi
\begin{flalign}
\label{H1122}
  & \mathcal{H}_{11} = [E_0+Af_1(k)]\openone+B(k_x\sigma_x + k_y\sigma_y)\notag\\
  & ~~~~~~~~+C(f_2(k)\sigma_x + f_3(k)\sigma_y),\\
  & \mathcal{H}_{22}=E'_0+A'f_1(k) , \\
  & \mathcal{H}_{12}=\mathcal{H}_{21}=0,
\end{flalign}
where we have defined functions $f_1(k)=k_x^2 + k_y^2$, $f_2(k)=k_y^2- k_x^2$, and $f_3(k)=2k_xk_y$.
Therefore, the diagonalization of parameter-dependent Hamiltonian (\ref{eq:Ham3*3})
results in the three-band energy spectra Eqs. (\ref{eq:SN1_eq1}-\ref{eq:SN1_eq2}) (see \ref{SN1}). 
We then performed a fitting procedure of these energy bands with first-principle data to obtain
the numerical values of parameters.
The comparison between DFT bands and the bands from our continuous model around the K point 
is shown in Supplementary Fig.\ref{fig:Al2B2-fit-K}.
One can see that around this point the fitted bands are in good agreement with the first-principles data. 
Furthermore, they also reproduce well the band dispersions 
in 2D BZ  (around K point), and therefore 
the locus of the touching points between these energy bands leads to the emergence of NL1 as depicted in Fig.~\ref{fig:Al2B2-KP-3D}(a).\\
%that reproduce very well the band dispersions in 2D BZ  and the locus of NL1 (see Figs. () and () ).\\
Having confirmed the existence of NL1 in Al$_2$B$_2$, we next study 
the reasons behind the formation of NL2, NL3, and NL4 which enclose
the $\Gamma$ point. In Fig.~\ref{fig:Al2B2-band-structure}(a), we observe that all of
these nodal lines are located completely below 
the Fermi level in the range of -2.35 to -1.35~eV which
can be revealed by ARPES measurements~\cite{zhou2018coexistence}.
Let us first focus on NL3 and NL4 that have distribution  near the the $\Gamma$ point. 
Here, two of the responsible bands are hole-like (labeled as $\lambda$ and $\delta$ bands) and the other
one is electron-like (labeled as $\nu$ band).
%\\\\\\\\\\\\\\\\\\\\\\\\\\\\\\\\\\\\\\\\\\\\\\
\begin{figure}[t!]
    \centering
    \includegraphics[width=.48\textwidth,valign=t]{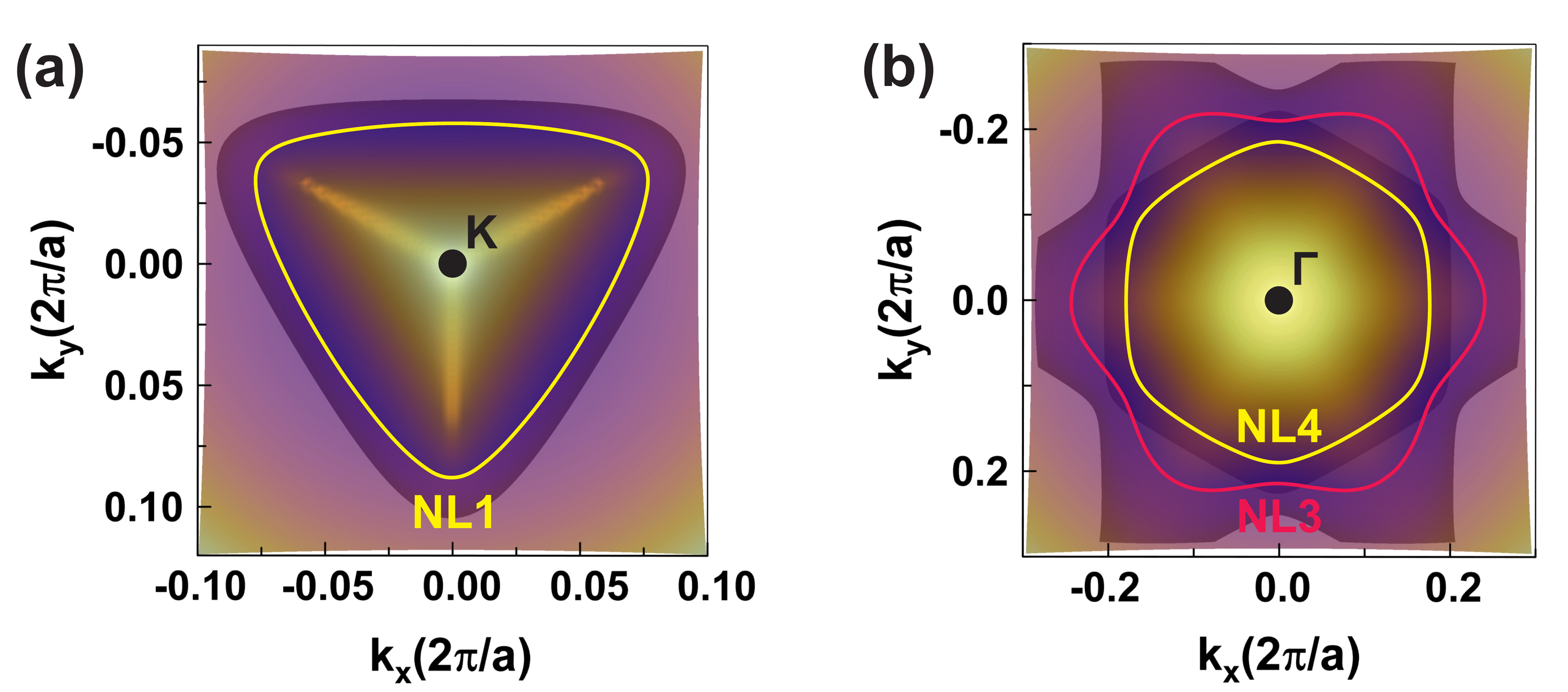}
    \caption{
             Transparent top views of crossing bands obtained
             by continuum model around (a) the K and (b)  $\Gamma$ points of Al$_2$B$_2$.
             The momentum distributions of gapless nodal points (NL1, NL2, and Nl3)
             are reproduced correctly by the continuum model.
            }
 \label{fig:Al2B2-KP-3D} 
\end{figure}
\iffalse, we label
the three highest responsible valence bands with , $\beta$, and $\gamma$ (see Fig.)
that their crossing would results in the formation of  NL1 and NL2. \fi
As seen from Figs.~\ref{fig:Al2B2-band-structure}(b) and (c), the $\lambda$ and $\delta$ bands 
are largely composed of B-p$_{x,y}$ orbitals while 
the $\nu$  band is mainly contributed by Al-p$_{z}$ and B-p$_{z}$ states.
The little group of wave vectors $\bm k$  along the $\Gamma$-M direction
with symmetry elements $E$, $C_2$, $M_{\sigma}$, and $M_h$
shares the same point group symmetry as $\Gamma$-K  and K-M paths. Therefore, the energy
bands of Al$_2$B$_2$ along this direction belong to one of the irreducible
representations of the $C_{2v}$ point group.
As shown in Fig.~\ref{fig:Al2B2_dft-wann-comparison}, a symmetry analysis unveils that along $\Gamma$-K ($\Gamma$-M) directions 
the $\lambda$,  $\delta$, and $\nu$ bands belong to three different 1D
irreducible representations $\Gamma_1$($\Gamma_2$), $\Gamma_2$($\Gamma_1$), and $\Gamma_4$, respectively.
For both symmetry directions, the corresponding bands with different IRs $\Gamma_1$ and $\Gamma_4$ have opposite
parities with respect to either $C_2$ or $M_h$ symmetry elements. Therefore, they will not hybridize with each 
other, allowing them to meet at NP3 and NP4$'$.
Similarly, for symmetry direction $\Gamma$-K ($\Gamma$-M), the energy
bands with IRs  $\Gamma_2$ and $\Gamma_4$ have opposite parities 
under mirror symmetries $M_h$ or $M_{\tau}$ ($M_{\sigma}$), leading to the formation of the nodal point NP4 (NP3$'$).
Here, a similar analysis to what we described for the formation
of NL1 could be applied to confirm that no gap opens  at the crossing between the corresponding bands along a typical
low-symmetry line X-$\Gamma$, leading to the emergence of NL3 and NL4.\\
We now turn to the method of invariants. This approach also provides a good description of 
momentum distributions of these nodal loops.
The little point group at $\Gamma$  point in the Al$_2$B$_2$  is $D_{6h}$.   At this 
point, the Bloch functions $\{ \psi^{\lambda},\psi^{\delta} \}$ and  $\psi^{\nu}$  transform according to the
IRs $\Gamma_6^+$ and $\Gamma_2^-$, respectively. Taking the set of symmetrical basis functions $\{ x-iy,x+iy \}$ and \{$z$\}, we obtained the basis 
matrices and the irreducible tensor components that transform accordingly as listed in Supplementary Table \ref{tab:Al2B2-G}.
Note that at the $\Gamma$ point we have $\bm k=-\bm k=0$. Therefore, according to the Herring test, these irreducible 
representations belong to the case '$a_1$'~\cite{bir1974symmetry}. On the other hand,  odd functions of $k_i^nk_j^m$ are not 
invariants of $\Gamma_{6}^{+} \otimes \Gamma_6^{+ \ast}$ and $\Gamma_2^{-} \otimes \Gamma_2^{- \ast}$.  As a
result, TR invariance implies no additional constraint.
Based on these two irreducible representations,  we
divide the $3 \times 3$ effective Hamiltonian $\mathcal{H}^\mathrm{\Gamma}$ into four blocks  and obtain 
each block using the method of invariants as follows
\iffalse
\begin{flalign}
\label{H1122}
  & \mathcal{H}_{11} = [E_0+A(k_x^2 + k_y^2)]\openone+B[(k_y^2- k_x^2)\sigma_x + 2k_xk_y\sigma_y]\notag\\
  & ~~~~~~~~+C[(k_y^4-6k_x^2k_y^2+ k_x^4)\sigma_x + 4(k_xk_y^3+k_x^3k_y)\sigma_y],\\
  & \mathcal{H}_{22}=E'_0+A'(k_x^2 + k_y^2) , \\
  & \mathcal{H}_{12}=\mathcal{H}_{21}=0,
\end{flalign}
\fi
\begin{flalign}
\label{H1122}
  & \mathcal{H}_{11} = [E_0+Af_1(k)]\openone+B[f_2(k)\sigma_x + f_3(k)\sigma_y]\notag\\
  & ~~~~~~~~+C[f_4(k)\sigma_x + f_5(k)\sigma_y],\\
  & \mathcal{H}_{22}=E'_0+A'f_1(k) , \\
  & \mathcal{H}_{12}=\mathcal{H}_{21}=0,
\end{flalign}
where $f_4(k)=k_y^4-6k_x^2k_y^2+ k_x^4$, and $f_5(k)=4(k_xk_y^3+k_x^3k_y)$.
The fitting procedure of the eigenvalues of this model 
with the DFT band structure is presented in  Supplementary Fig.~\ref{fig:Al2B2_fit_G}.
Using the fitted values we have shown in Fig.~\ref{fig:Al2B2-KP-3D}(b) a transparent top view of
energy bands corresponding to NL3 and Nl4, respectively.  
As seen, our
effective Hamiltonian reproduces correctly the hexagon (NL3) and hexagram (NL4) form of the momentum distribution of gapless nodal points.\\
The other nodal line that we shall examine is NL2.
Along the $\Gamma$-M direction, the IRs of touching bands at the NP2$'$ point are identical to 
those of NP4$'$. Hence, the same group theory analysis is applied to explain the formation of topological nodal point NP2$'$.
On the other hand,  along the $\Gamma$-K direction, the NP2 nodal point  emerges 
due to the crossing of two energy bands with different IRs $\Gamma_1$ and $\Gamma_3$. 
Here, the protection of this point is justified by the opposite parity of the two bands under $M_h$ or $M_{\tau}$.
It should be noted that the same interpretation as discussed earlier applies to other touching points of NL2  whose point groups are $C_s$.\\
The remaining nodal line in Al$_2$B$_2$  is NL5 which encloses the M point.
This nodal loop is located at least 0.48~eV above the Fermi energy and is less significant.
For this nodal line, along both $\Gamma$-M and $\Gamma$-K directions
the corresponding bands of touching points NP5 an NP5$'$
have the same IRs $\Gamma_1$ and $\Gamma_4$.  As the wave functions of these bands have opposite 
parities under $C_2$ and $M_h$, they prohibit the interaction between them which allow the appearance  of these
nodal points.
\iffalse{
As shown in Fig., a symmetry analysis unveils that along both $\Gamma$-M  and  $\Gamma$-K directions
the $\lambda$,  $\delta$, and $\nu$  bands belong to three different 1D 
irreducible representations $\Gamma_2$, $\Gamma_1$, and $\Gamma_4$, respectively.
As bands $\delta$ and $\nu$ with different IRs  $\Gamma_1$ and $\Gamma_4$ have opposite 
parities with respect to either $C_2$ or $M_h$ symmetries, they will not hybridize with each
other, allowing them to meet at NP3 and NP3$'$. Also,  there is no gap opening at the crossing between $\lambda$ 
and $\nu$ bands along $\Gamma$-K ($\Gamma$-M), since they have opposite parities under the mirror
symmetries $M_h$ or $M_{\tau}$ ($M_{\sigma}$), forming the nodal point NP4 (NP4$'$).
}
////////////////////////////////////\\
Since we have ignored the SOC, the Mz symmetry is  
respected in this configuration.

Figure .. illustrates the  3D  energy bands near the K point for the two bands whose crossing results into the 
formation of NL1.\fi
\\
\\
\textbf{DNLs in AlB$_4$.}
Now that the electronic and topological properties of  monolayer Al$_2$B$_2$ were investigated, we move 
to examine the case of monolayer AlB$_4$.  This configuration share the same symmetry
as Al$_2$B$_2$ with $D_{6h}$ point group. Figure \ref{fig:mirror-planes}(b) depicts those  symmetry elements of AlB$_4$ which  
we are interested in the following. 
The band structure of AlB$_4$ along the highly symmetric
paths (M-$\Gamma$-K-M) is displayed in Fig.~\ref{fig:AlB4-band-structure}(a).
One can observe from Figs.~\ref{fig:AlB4-band-structure}(b) and (c) that around the $\Gamma$ point and 
close to the Fermi level energy bands 
mainly originate from boron $\sigma$ orbitals (B-s and  B-p$_{x,y}$).  However, 
there exist also pure $\pi$-bonding boron states as well as mixed orbitals
(constructed from B-p, B-s, and Al-s)  that 
\begin{figure}[t!]
    \centering
    \hspace{-0.5 cm}
    \includegraphics[width=.48\textwidth,valign=t]{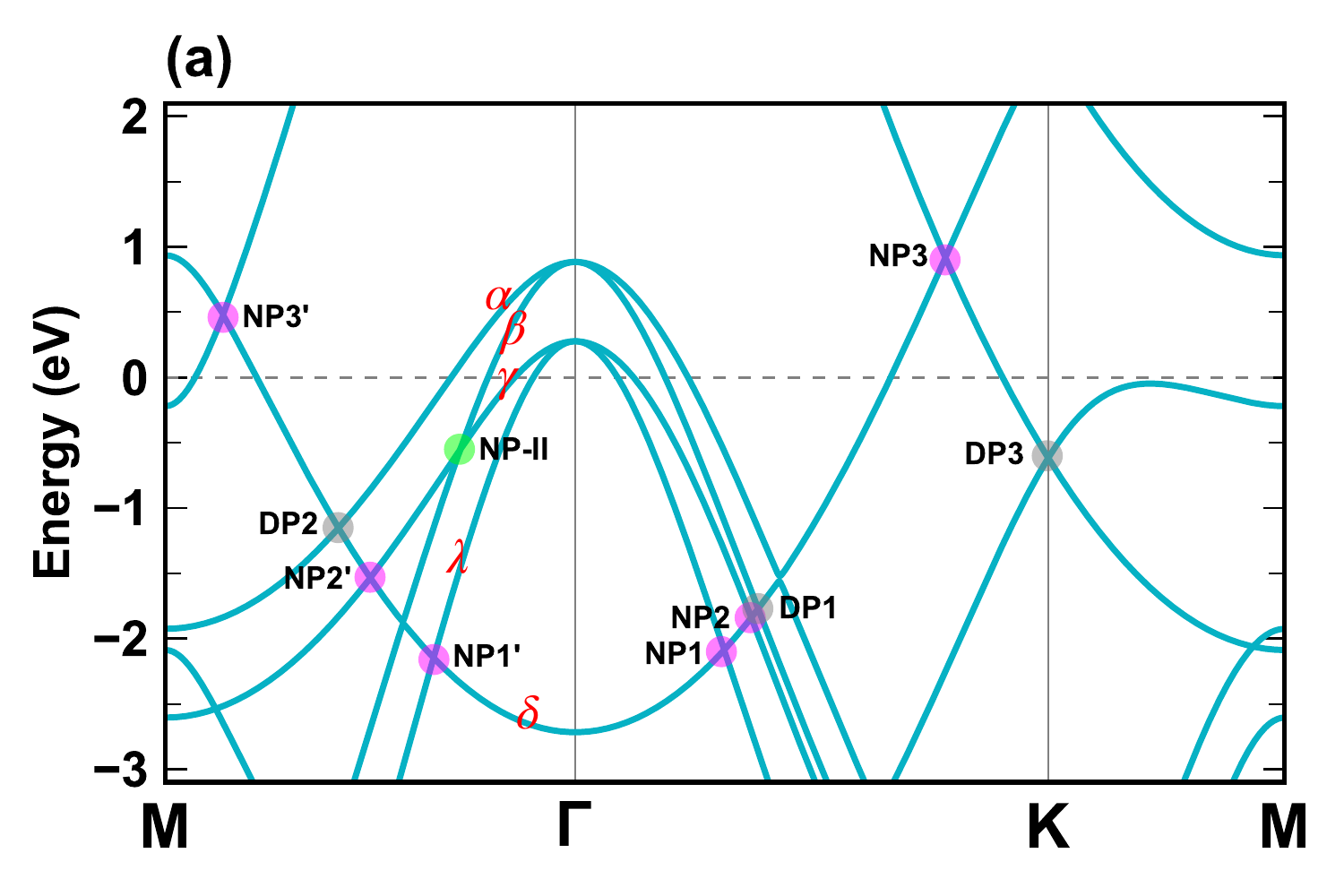}  \\ %\quad
    \hspace{0.1 cm}
    \includegraphics[width=.235\textwidth,valign=t]{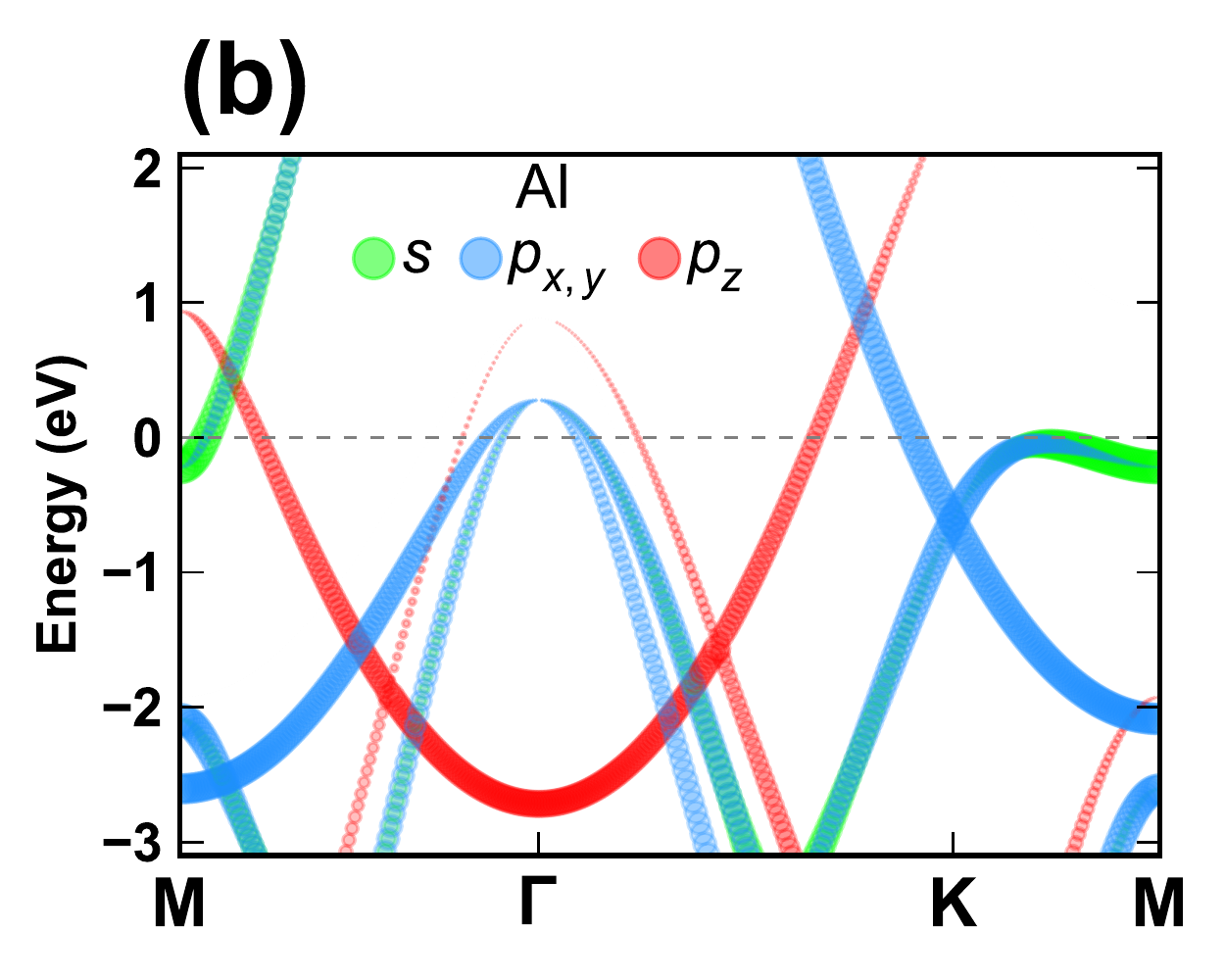} %\quad
    \hspace{-0.2 cm}
    \includegraphics[width=.235\textwidth,valign=t]{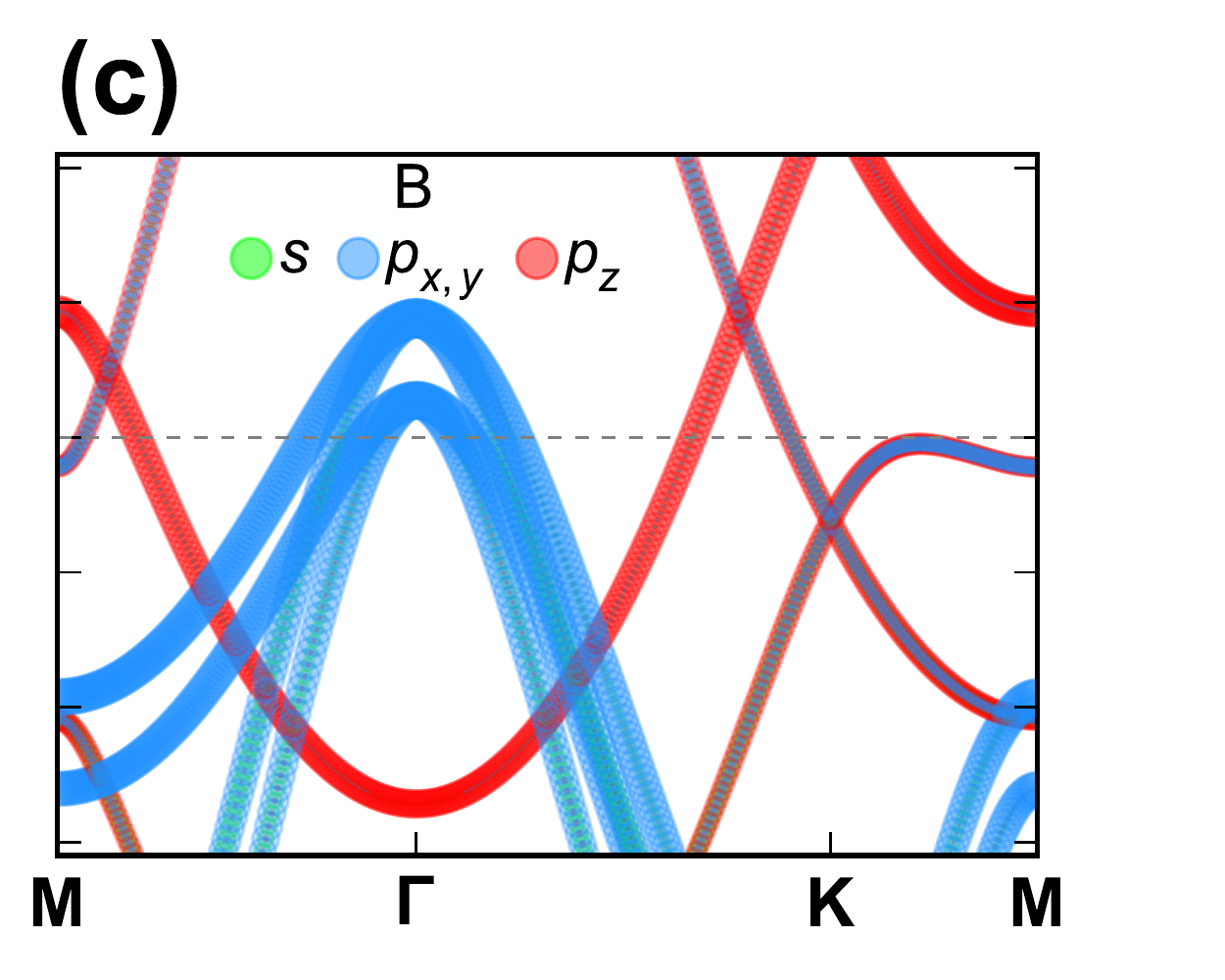} %\quad
    \caption{
             (a) The  calculated  electronic  band structure  of Al$_2$B$_2$ configuration.  Energy  dispersion 
             display multiple band crossing features including both Dirac and nodal points as presented by magenta circles.
             Special bands are marked by Greek letters.
             (b) and (c) Atom-projected band structures for Al$_2$B$_2$ contributed by Al and B atoms, respectively.  
             The size of dots is proportional to the weight of contributed orbitals.
            }
 \label{fig:AlB4-band-structure} 
\end{figure}
\begin{figure}[h!]
    \centering
    %\hspace{0.1 cm}
    %\includegraphics[width=.4\textwidth,valign=t]{img/Al2B2_DFT_WANN_compare_bands.pdf} \\ %\quad
    %\hspace{-0.2 cm}
    \includegraphics[width=.48\textwidth,valign=t]{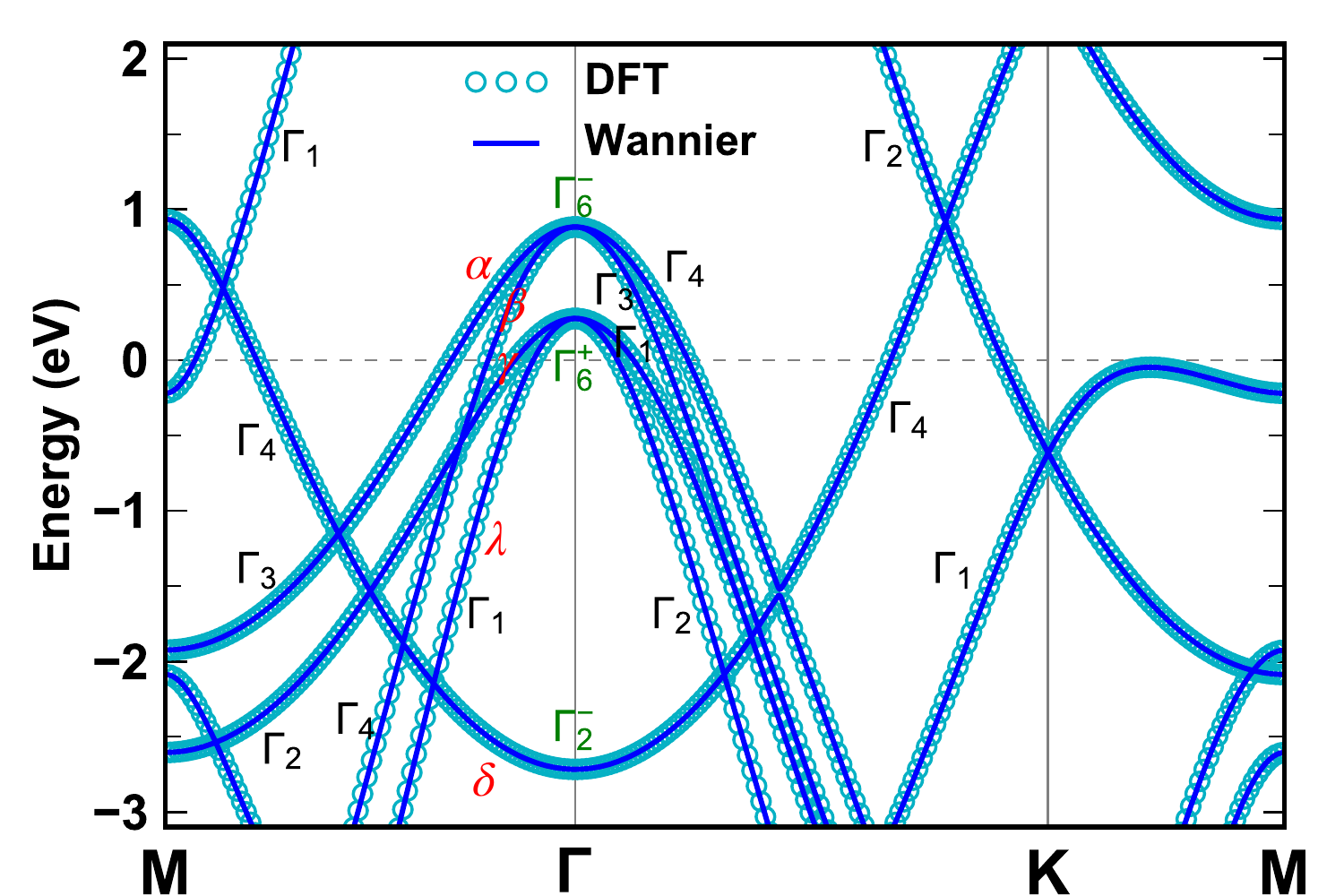} %\quad 
    \caption{ 
             Comparison of the band structures of AlB$_4$ as calculated by DFT and the WTB Hamiltonian. 
             The Wannier band structure exactly matches the DFT results.
            } 
 \label{fig:AlB4_dft-wann-comparison} 
\end{figure}
\begin{figure*}[t!]
    \centering
    \includegraphics[width=.65\textwidth,valign=t]{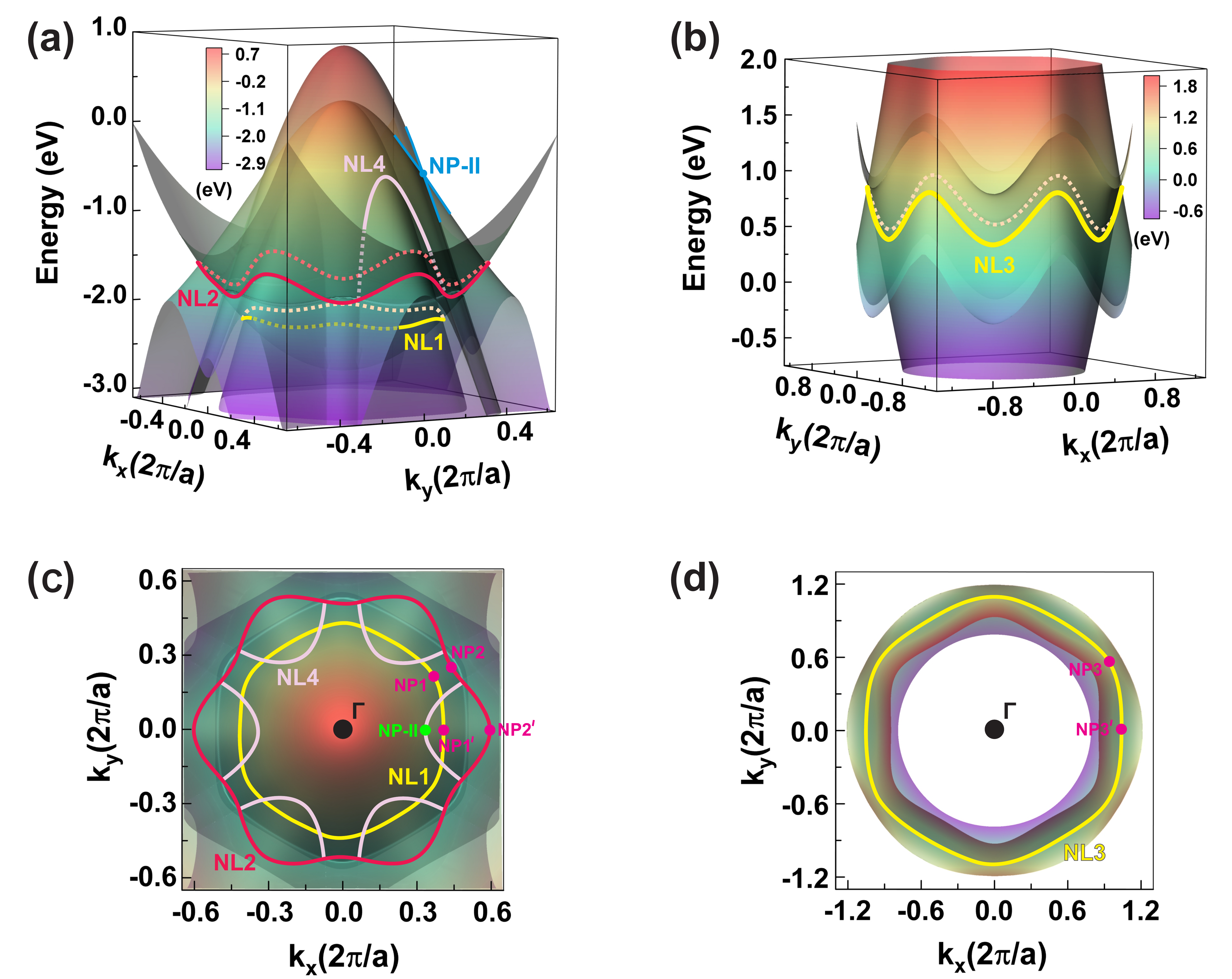} %\quad
    \caption{
             (a) and (b) 3D visualizations of energy bands obtained from  AlB$_4$  
             WTB Hamiltonian. Bands whose crossings lead to the formation of nodal lines around
             the $\Gamma$ point are shown. The extensions of nodal points can be seen vividly
             throughout 3D energy-momentum spaces which form the characteristic topology of each NL.
             Color bars represent the energy of electrons.
             (c) and (d) The transparent top views of (a) and (b). The momentum distributions of gapless nodal points are shown. 
             Note that in Figs.(b) and (d) some data values were removed for better visualization.
            } 
 \label{fig:AlB4-3d-plots} 
\end{figure*}cross the Fermi level. In a recent study, the role of these low-energy bands  in determining 
the superconducting behavior of the system has been explored~\cite{zhao2019two}.
It was revealed that the AlB$_4$ film captures a robust three-gap superconducting nature with a high critical temperature  ($\sim$47~K)
which provides a platform for further investigation of multigap superconductors.
Surprisingly, we found that along with the superconducting behavior of these energy bands
there are linear band  crossing features 
%(including nodal points (NPs) and Dirac points (DPs))
that emerge as a consequence of 
lattice symmetries. This led us to investigate the topological aspects 
of the AlB$_4$ band structure which contains several dispersive NLs and Dirac cones.\\
Let us label the mentioned bands using Greek letters as indicated in Fig.~\ref{fig:AlB4-band-structure}(a).
In the energy range from -2.2 to 1~eV, the crossing points consist of either nodal points (NPs) or Dirac points (DPs).
There are three Dirac points as indicated by DP$1$ to DP$3$. 
Also, the nodal points  NP$i$, and NP$i'$ 
($i$ runs from 1 to 3) are pertinent points which belong to the 2D nodal loop NL$i$. 
In addition, there exists a distinct nodal point (NP-II) that corresponds
to an open  2D type-II nodal line (NL4).  This is to the best of our knowledge the first evidence 
for the existence of a 2D nonmagnetic open type-II NL in systems at weak SOC limit.
Figure~\ref{fig:AlB4_dft-wann-comparison} shows the 
comparison of Wannier band structure of AlB$_4$ with
DFT data. Using the obtained WTB Hamiltonian we provide in Fig.~\ref{fig:AlB4-3d-plots}
3D energy bands of AlB$_4$ near the Fermi level in the vicinity of the $\Gamma$ point.
An inspection of these 3D bands and the corresponding 
transparent top views (Figs.~\ref{fig:AlB4-3d-plots}(c) and (d)) reveals that the momentum distribution of
type-I nodal loops NL1 and NL3  show a hexagon, while for the other 
type-I nodal loop i.e. NL2, it is a hexagram. Besides, one can see from Figs.~\ref{fig:AlB4-3d-plots}(a) and (c) that the 
open type-II nodal lines  NL4 are terminated to the closed nodal hexagram NL2.
%The coexistence of type-I and type-II  NLs in this structure 
%can be confirmed through ARPES~\cite{feng2017experimental} or infrared spectroscopy~\cite{shao2020electronic} measurements, which are
%two ideal probes of the nodal line dispersion in NLSMs.
The coexistence of these two nodal line types which can be confirmed through ARPES~\cite{feng2017experimental} or 
infrared spectroscopy~\cite{shao2020electronic} measurements,  may provide an ideal platform to study many novel 
physical properties~\cite{yu2016predicted,udagawa2016field,o2016magnetic} that have been
proposed for NLSMs. For example, it is preferred
to examine the signatures of Landau levels spectrum collapse~\cite{yu2016predicted} on
a 2D platform rather than in a 3D  material candidate.
Also, it  might be beneficial to utilize this topological
material to explore experimentally the anticipated enhanced correlations in nodal line semimetals~\cite{shao2020electronic}.
Furthermore, the interplay of the superconductivity and these topological Dirac nodal line 
states suggests that topological superconducting phases~\cite{jin2019topological} may be 
realized in this structure which is a subject for further investigations.\\
The hexagon and hexagram shapes of NL1 (NL3) and NL2
are due to the fact that the underlying symmetries and irreducible representations
of the corresponding bands are identical to that of NL4 and NL3 in Al$_2$B$_2$ lattice, respectively.
Therefore, the same group theory and method of invariants analysis 
applies to explain the formation mechanism of these NLs, and no
further discussion seems necessary to highlight here.\\
Before we turn to the open nodal loop NL4,  we shall briefly explain
the reasons why  Dirac points DP1, DP2, and DP3  occur in the band structure of this configuration.
DP1 and DP2 locate along the  $\Gamma$-K and  $\Gamma$-M directions $\sim$1.18~eV and $\sim$1.77 eV below the Fermi level, respectively. 
Since the number of such points in the star of $k$ in the BZ is six,
one expect to see the same number of Dirac points.
As seen in Fig.~\ref{fig:AlB4_dft-wann-comparison},  the IRs of the touching bands around both Dirac points 
DP1 and DP2 are $\Gamma_3$ and $\Gamma_4$.   Therefore, just similar to the case of DP1 in Al$_2$B$_2$,
the opposite parities of $C_2$ rotation or mirror $M_\sigma$ (or $M_\tau$) allow the two bands 
to touch each other at DP1 and DP2.  It is worth mentioning that at first glance, 
one may expect that similar to a typical point NP$i$, these points also belong to a nodal line distribution.
However, the situation is different here. Along a typical low symmetry X-$\Gamma$ direction  with point group $C_s$, both bands (around DP1 and DP2) have
the same IR $\Gamma_2$ which prevents them to touch each other, leading to the opening of a gap. 
The other Dirac point that locates at the K point and $\sim$0.65~eV below the Fermi level is DP3 which has the characteristics of the Dirac points in graphene.\\
Now, we will focus on the topological 
aspects of type-II nodal line NL4. Along the $\Gamma$-M direction, the two electron-like bands $\beta$ and $\gamma$ (with the same slope sign) meet
$\sim$0.6 eV below the Fermi level and show a band crossing as labeled by NP-II (Fig.~\ref{fig:AlB4-band-structure}(a)). 
Figures \ref{fig:AlB4_BS_XG}(a) and (b) display band structures of AlB$_4$ along two typical
low-symmetry lines X-$\Gamma$ where X is a point along the high-symmetry M-K direction (Fig.~\ref{fig:AlB4_BS_XG}(c)). 
As seen, for a typical line X1-$\Gamma$, where X1 is between K and C, 
there exists no crossing between the $\beta$ and $\gamma$ bands.
Conversely, when X is between C and M, these two bands crosses along 
the corresponding X-$\Gamma$ path (Fig.11b).
Therefore, the crossing point of the $\gamma$ and $\beta$ bands are only distributed
in certain Brillouin zone slices, shown in Fig., to form a set of open type-II nodal lines (NL4) in the system.
Here, C specifies the low symmetry path wherein NL4 is linked with NL2 (see Fig.~\ref{fig:AlB4-3d-plots}(c)).
%band crossings for all $k$ points between  M and X2 are formed which disperse  over a rather large energy range of .
The rather large energy dispersion of these type-II nodal lines ($\sim$1.3~eV) makes AlB$_4$ 
a distinguished 2D candidate for realization of novel topological properties.
\begin{figure}[t!]
    \centering
    \includegraphics[width=.46\textwidth,valign=t]{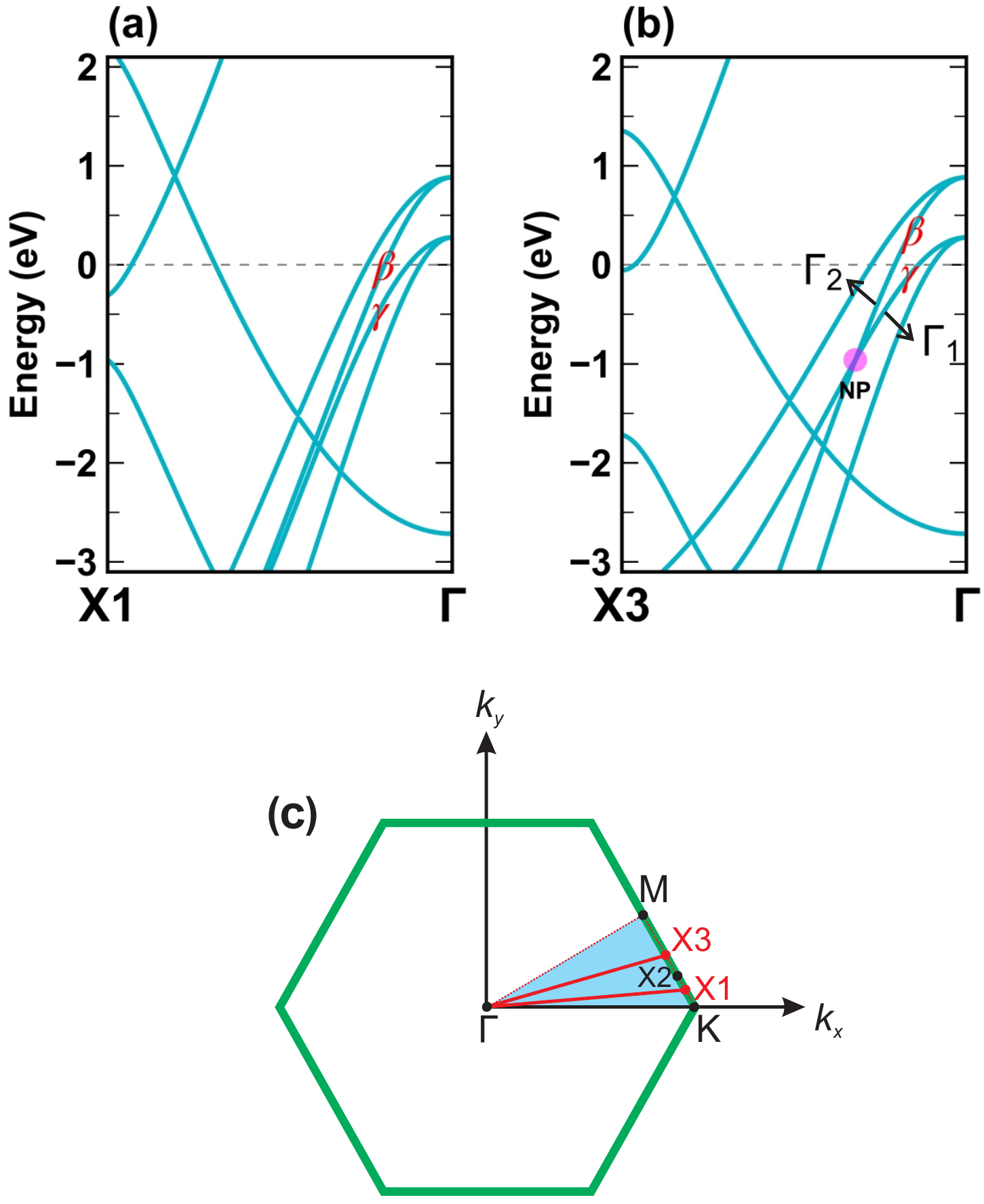}
    \caption{ 
             (a) and (b) Electronic band structures of AlB$_4$ along typical
             low-symmetry paths X1-$\Gamma$ and X3-$\Gamma$ in the BZ as shown by rel lines in (c).
             The crossing of $\beta$ and $\gamma$ bands exists for a typical path X-$\Gamma$ where X 
             is located between X2 and M points. 
            } 
 \label{fig:AlB4_BS_XG} 
\end{figure}\\
The formation of the topologically protected momentum distribution
NL4 is justified as follow. Along the high symmetry direction $\Gamma$-M
with point group $C_{2v}$, the IRs of $\beta$ and $\gamma$ bands are $\Gamma_4$ and $\Gamma_2$, respectively 
(see Fig.~\ref{fig:AlB4_dft-wann-comparison}).
The opposite parities of these two bands under mirror symmetries $M_h$ or $M_{\sigma}$ decouple 
the associated electronic states and allow the formation of the nodal point NP-II.
Moreover, along the typical low symmetry line X3-$\Gamma$ with point group $C_{s}$, these bands have different IRs $\Gamma_2$ and $\Gamma_1$. The situation 
is just similar to the case of NL1 in Al$_2$B$_2$  which permits 
the appearance of open type-II NL4.\\
Eventually,  using the method of invariants, we derive
the effective  Hamiltonian model which describes quite well those nodal lines of AlB$_4$ that locate below the Fermi level.
For the little group $D_{6h}$ at $\Gamma$ point the
relevant Bloch states belong to the doubly degenerate IRs $\Gamma_{6}^{-}$, $\Gamma_{6}^{+}$, and 
nondegenerate IR $\Gamma_{2}^{-}$ as shown in Fig.~\ref{fig:AlB4_dft-wann-comparison}.
Thus, the corresponding Bloch functions allow us to construct an effective  $5\times5$ block diagonal 
Hamiltonian $\mathcal{H}^\mathrm{\Gamma}$ around the $\Gamma$ point as
\begin{equation}
  \label{eq:dham}
  \mathcal{H}^\mathrm{\Gamma} = \left( \begin{array}{ccc}
   \mathcal{H}_{11} & \mathcal{H}_{12} & \mathcal{H}_{13}\\
   \mathcal{H}_{21} & \mathcal{H}_{22} & \mathcal{H}_{23}\\
   \mathcal{H}_{31} & \mathcal{H}_{32} & \mathcal{H}_{33}
    \end{array}\right) \, .
\end{equation}
Therefore, using the method of invariants 
we adopt $\{ x-iy,x+iy \}$, $\{ x-iy,x+iy \}$, and \{$z$\} as basis functions to 
obtain the basis matrices as well as tensor components (see \ref{SN3}).  Inserting the obtained 
basis matrices and tensor components (TR adds no additional constraint 
as discussed in the case of Al$_2$B$_2$) in each block of $\mathcal{H}^\mathrm{\Gamma}$ we arrive at
\iffalse
\begin{flalign}
\label{H112233}
  & \mathcal{H}_{11} = [E_0+A(k_x^2 + k_y^2)]\openone+B[(k_y^2- k_x^2)\sigma_x + 2k_xk_y\sigma_y]\notag\\
  & ~~~~~~~~+C[(k_y^4-6k_x^2k_y^2+ k_x^4)\sigma_x + 4(k_xk_y^3+k_x^3k_y)\sigma_y],\\
  & \mathcal{H}_{22} = [E'_0+A'(k_x^2 + k_y^2)]\openone+B'[(k_y^2- k_x^2)\sigma_x + 2k_xk_y\sigma_y]\notag\\
  & ~~~~~~~~+C'[(k_y^4-6k_x^2k_y^2+ k_x^4)\sigma_x + 4(k_xk_y^3+k_x^3k_y)\sigma_y],\\ 
  & \mathcal{H}_{33}=E''_0+A''(k_x^2 + k_y^2) , \\
  & \mathcal{H}_{12}=\mathcal{H}_{21}=0. \\
  & \mathcal{H}_{ij}=0 ~~~{fto} ~~~ {i \neq j}.
\end{flalign}
\fi
\begin{flalign}
\label{H112233}
  & \mathcal{H}_{11} = [E_0+Af_1(k)]\openone+B[f_2(k)\sigma_x + f_3(k)\sigma_y]\notag\\
  & ~~~~~~~~+C[f_4(k)\sigma_x + f_5(k)\sigma_y],\\
  & \mathcal{H}_{22} = [E'_0+A'f_1(k)]\openone+B'[f_2(k)\sigma_x + f_3(k)\sigma_y]\notag\\
  & ~~~~~~~~+C'[f_4(k)\sigma_x + f_5(k)\sigma_y],\\ 
  & \mathcal{H}_{33}=E''_0+A''f_1(k) , \\
  & \mathcal{H}_{ij}=0 ~~~ ~~~ {i \neq j},
\end{flalign}
where we have previously defined $f_i(k)$ functions. By fitting the parameters with DFT 
results (see Supplementary Fig.~\ref{fig:AlB4-fit-G}),
our continuous five-band model describes very well the formation of open type-II
nodal line NL4 as well as NL1 and NL2 around the $\Gamma$ point as shown in Figs.\ref{fig:AlB4-KP-3D}(a) and (b).
\begin{figure}[t!]
    \centering
    \includegraphics[width=.48\textwidth,valign=t]{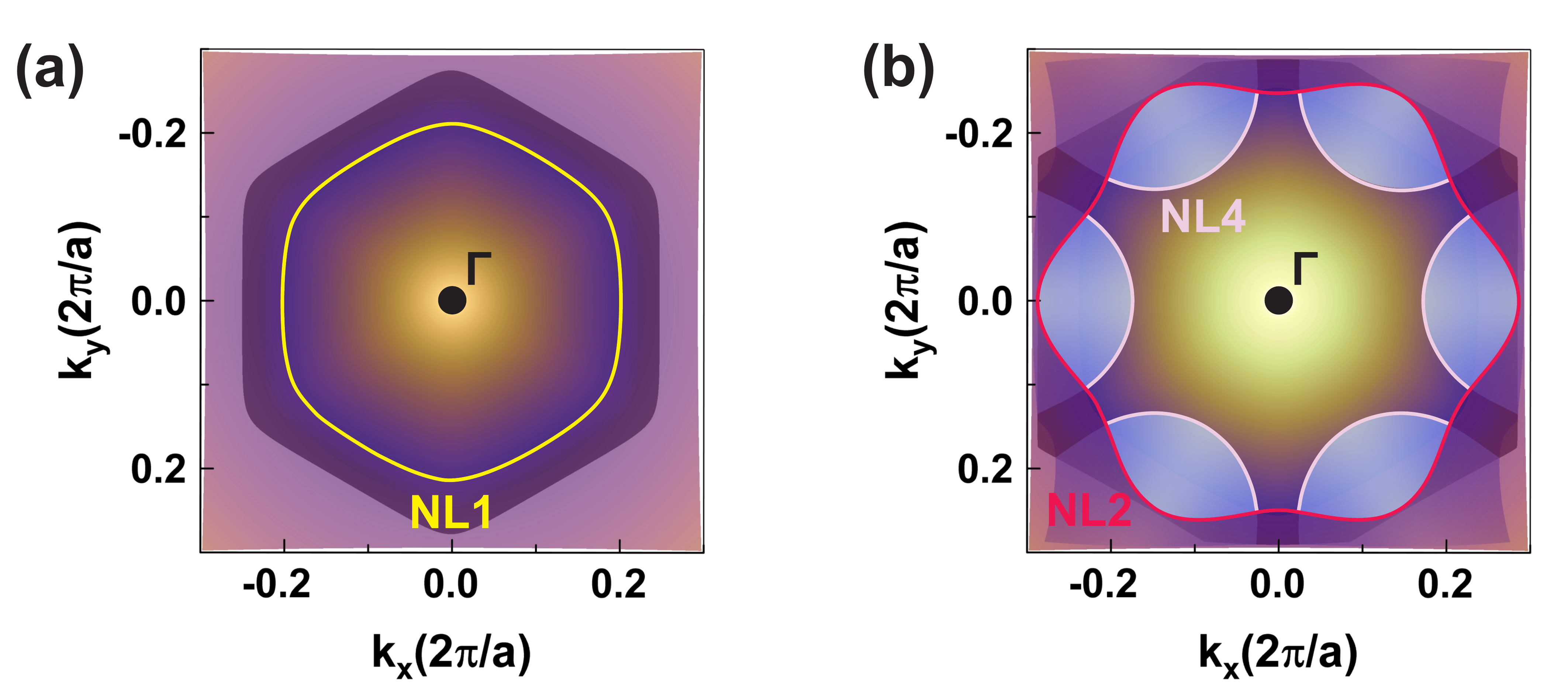}
    \caption{
             Transparent top views of crossing bands obtained
             by continuum model around the $\Gamma$ point of AlB$_4$. 
             The momentum distributions of gapless nodal lines (a) NL1 
             and (b) NL2 and NL4 are reproduced correctly by the continuum model.
            }
 \label{fig:AlB4-KP-3D} 
\end{figure}
\iffalse 3D inspection of the band structure in the whole BZ  reveals that in the energy range from -3.5 to 2 eV
there exist three independent type-I nodal loops and a set of open type-II
nodal lines terminated to the lobes of a closed nodal hexagon.\fi
%Such a nodal line metal may provide an excellent platform to investigate correlation effects because of linearly dispersing Dirac quasiparticles in its spectrum.
\\Finally, we make some notes on the freestanding monolayer AlB$_2$. 
Although, group theory analysis of the energy bands unveils that there exist
some topologically protected Dirac points along the high symmetry lines 
in the BZ of this structure,
the absence of the in-plane mirror symmetry 
reduces the point group of the system to $C_{6v}$ and 
thus prevents formation of any nodal line feature
in the momentum space (See \ref{SN4}).  
It is worth mentioning that the monolayer AlB$_2$ 
with a B-terminated surface has been recently synthesized on Al(111)~\cite{geng2020}
and the existence of the p-type Dirac Fermions in this configuration was 
confirmed by the recent experimental ARPES measurements~\cite{geng2020}. This may 
serve as a possible way to explore our predicted topological NLSM states in 
aluminum diborides thin films with in-plane mirrors.
\iffalse
.....Because the SOC
in GdAg 2 is rather weak, the energy gap is negligible and
therefore NL3 is approximately preserved.

As we noted, due to the very weak SOC
in these structures, the energy gap is negligible and
therefore NL3 is approximately preserved.
 
  orbital-characteristics
  
  It should be noted here that these
stable monolayer and trilayer films have not been reported
experimentally to date.

Let us now turn to the mechanism of emergence of Dirac points.
For convenience,  in the vicinity of the mentioned nodal and Dirac 
points, we label the electronic bands using greek letters as indicated in Fig. 
As seen, along the Γ-K direction the crossing of .. and .. bands leads to 
DP1. The little group for Al2B2 configuration along
this high symmetry direction is C2v with two perpendicular 
mirror reflections planes; Mσ and Mh which their intersection introduces 
the twofold rotation axis C2.  The ...and .. bands have opposite parities with respect to C2 rotation as... in ...

The hexagonal diborides have / share the same centrosymmetric crystal structure
with the space group P6/mmm (No.191). As shown in Fig.\ref{Structure_BZ} (a), the 
boron layer has graphene-like honeycomb lattice while metal atoms have close-packed hexagonal structure.
The positions of boron Atoms are (1/3,2/3,1/2) and (2/3,1/3,1/2) and the metal atoms locating at (0,0,0).
\fi
\\
\\
\textbf{\large {Summary and conclusions}}\\
In summary, using first-principles calculations and group theory analyses, we accurately investigated 
the electronic structure of Al$_2$B$_2$ and AlB$_4$ thin films to 
reveal the existence of many nontrivial topological features in these systems.
The dynamical stability of these structures 
was also proven by calculation of their phonon spectra.
Our results indicate that the electronic band structures of both configurations host multiple 
topologically protected band crossings including 2D Dirac nodal lines
as well as Dirac points.
We then used symmetry analyses to explain the protection mechanism of the nodal lines 
and then applied the method of invariants to construct proper effective Hamiltonians 
for systematic investigation of the formation mechanism of these nodal lines.
We demonstrated formation of five and three type-I topological NLs in the low energy electronic states of 
the Al$_2$B$_2$ and AlB$_4$ compounds, respectively.
It is argued that all of the nodal line states are protected by 
the time reversal and in-plane mirror symmetries.
In the case of Al$_2$B$_2$, three concentric nodal loops sit below the Fermi level and surround the $\Gamma$ point,
while an isolated nodal ring appears around the M point above the Fermi level.
Moreover, it hosts a dispersive nodal loop centered around 
the K point, appearing as six arcs in the corners of the Brillouin zone. 
This nodal loop crosses the Fermi level with a considerable dispersion and thus 
provide an excellent platform to study the fascinating characteristics of dispersive Dirac nodal lines.
For the AlB$_4$ thin film, on the other hand, three concentric nodal loops were evidenced 
in the low energy electronic states around the $\Gamma$ point 
with hexagon or hexagram momentum distributions. 
More importantly, we identify the first evidence for the emergence of a set of open 2D nonmagnetic 
type-II nodal arcs in the AlB$_4$ Brillouin zone,
linked with a type-I nodal line. 
The rather large energy dispersion ($\sim$1.3 eV) of these Dirac nodal arcs, 
their coexistence with multiple type-I NLs, and
the high-temperature superconducting behavior of AlB$_4$
suggest this structure as a distinguished material 
for advanced research on 2D topological superconductors.
\\
\\
\textbf{\large {Methods}}\\
All structural and electronic properties are obtained within the Kohn-Sham DFT calculations with 
the full potential linearized augmented plane wave (LAPW) method as implemented in the 
computer package WIEN2k~\cite{WIEN2k191}.
The generalized gradient approximation (GGA) in the Perdew-Burke-Ernzerhof (PBE) 
formulation~\cite{Perdew1996Generalized,Blochl1994Projector} was used 
for the exchange-correlation functional.
The expansion cutoff $R_{\mathrm{MT}}K_{\mathrm{max}}$ was set to 8, while a $30\times30\times1$ $\Gamma$-centered
$k$ mesh was applied to sample the BZ for self-consistent-field calculations.
The crystal structures are fully relaxed with a total energy and force convergence 
criteria of $10^{-4}$ Ry and $10^{-3}$  Ry/a.u., respectively.
A vacuum of more than 20~{\AA} was used to minimize the interactions between the neighboring replica of the system.
The maximally localized Wannier functions were constructed by employing
the Wannier90 package~\cite{Pizzi2020Wannier90}.
The Al s, p and B s, p orbitals are chosen as initial projections for WTB model construction.

In order to calculated the phonon spectra, the interatomic force constants were obtained 
through a supercell approach by using the HIPHIVE~\cite{Hiphive} package.
We used $10\times10\times1$ supercells and then generated the reference rattled structures
by applying displacements, randomly sampled from a normal distribution with a standard deviation of 0.01~{\AA}.
The HIPHIVE package employs VASP~\cite{kres1,kres2} for the required DFT calculations,
being performed with the projector augmented-wave (PAW)~\cite{PhysRevB.50.17953}
pseudopotentials and PBE exchange-correlation functional. 
The PAW plane-wave energy cutoff was set to 500 eV.
After finding the force constants, 
the phonon frequencies were calculated using the PHONOPY package~\cite{phonopy}.
%\section{ACKNOWLEDGMENTS}
\bibliography{References}
~\\
\textbf{\large{Author contributions}}\\
S. Abedi and E. Taghizadeh Sisakht contributed equally to this work.
\\
\\
\textbf{\large{Conflicts of interest}}\\
There are no conflicts to declare.
\\
\\
%\textbf{\large{Acknowledgements}}\\
%/////////////////////////////////////
%//   Supplementary Information     //
%/////////////////////////////////////
\onecolumngrid
\newpage
\setcounter{section}{0}
\setcounter{equation}{0}
\setcounter{figure}{0}

\renewcommand {\thesubsection} {Supplementary Note \arabic{subsection}}
\renewcommand{\theequation}{S\arabic{equation}}
\renewcommand{\thetable}{S\arabic{table}}
\renewcommand{\thefigure}{S\arabic{figure}}
\renewcommand{\theHfigure}{A\arabic{figure}}
\renewcommand{\figurename}{Supplementary Fig}

\iffalse
\renewcommand{\thesection}{\Alph{section}}
\renewcommand{\thefigure}{S\arabic{figure}}
\renewcommand{\thetable}{S\Roman{table}}
\newcommand*{\citenamefont}[1]{#1}
\newcommand*{\bibnamefont}[1]{#1}
\newcommand*{\bibfnamefont}[1]{#1}
%\newcommand*{\bibinfo}[1]{#1}
%\newcommand*{\bibfield}[1]{#1}
%\renewcommand{\thesubsection}{SI\arabic{subsection}}
\fi
\section*{Supplementary Information}
%\subsection{Supplementary Note 1\label{SN1}}
\subsection{Supporting information of the continuum model for Al$_2$B$_2$ configuration around the K point.\label{SN1}}
\begin{table}[h!]
  \caption[]{
             Symmetrized matrices for the invariant
             expansion of the diagonal blocks $\mathcal{H}_{ii}$ for the point group $D_{3h}$.
            }
$\arraycolsep 0.8em
\renewcommand{\arraystretch}{1.2}
\begin{array}{clclcl} \hline \hline
\mbox{Block} &
\multicolumn{1}{l}{\mbox{Representations}} &
\multicolumn{2}{l}{\mbox{Symmetrized matrices}} &
\multicolumn{2}{l}{\mbox{Tensor components}} \\ \hline

\mathcal{H}_{11} & \Gamma_6 \otimes \Gamma_6^\ast    & \Gamma_1: & \openone            & \Gamma_1: & 1; \; {k_x^2 + k_y^2} \\
                 & = \Gamma_1 + \Gamma_2 + \Gamma_6  & \Gamma_2: & \sigma_z            & \Gamma_2: & -  \\
                 &                                   & \Gamma_6: & \sigma_x, \sigma_y  & \Gamma_6: & (k_x, k_y); (k_y^2- k_x^2, 2k_x k_y)\\[1ex]
  
\mathcal{H}_{22} & \Gamma_4 \otimes \Gamma_4^\ast    & \Gamma_1: & \openone            & \Gamma_1: & 1; \; {k_x^2 + k_y^2} \\
                 & = \Gamma_1 \\ \hline \hline

\end{array}$
\label{tab:Al2B2-K}
\end{table}

\begin{equation}
E_{1,2}(k_x,k_y)={E_0}+A ({k_x}^2+{k_y}^2)\pm \sqrt{B^2 ({k_x}^2+{k_y}^2)+C^2 ({k_x}^2+{k_y}^2)^2-2BC( {k_x}^3 - 3{k_x} {k_y}^2)},
\label{eq:SN1_eq1}
\end{equation}
\begin{equation}
E_{3}(k_x,k_y)={E'_0}+A'( {k_x}^2+ {k_y}^2).
\label{eq:SN1_eq2}
\end{equation}
The fitted parameters are $E_0=1.19$ eV, $A=-113.16$ eV\AA$^2$, $B=8.78$ eV\AA,  $C=132.10$ eV\AA$^2$, $E'_0=-0.56$ eV, $A'=66.33$ eV\AA$^2$ .
The fitted bands are shown in Supplementary Fig.~\ref{fig:Al2B2-fit-K}.
\begin{figure}[h!]
  \includegraphics[width=0.50\columnwidth]{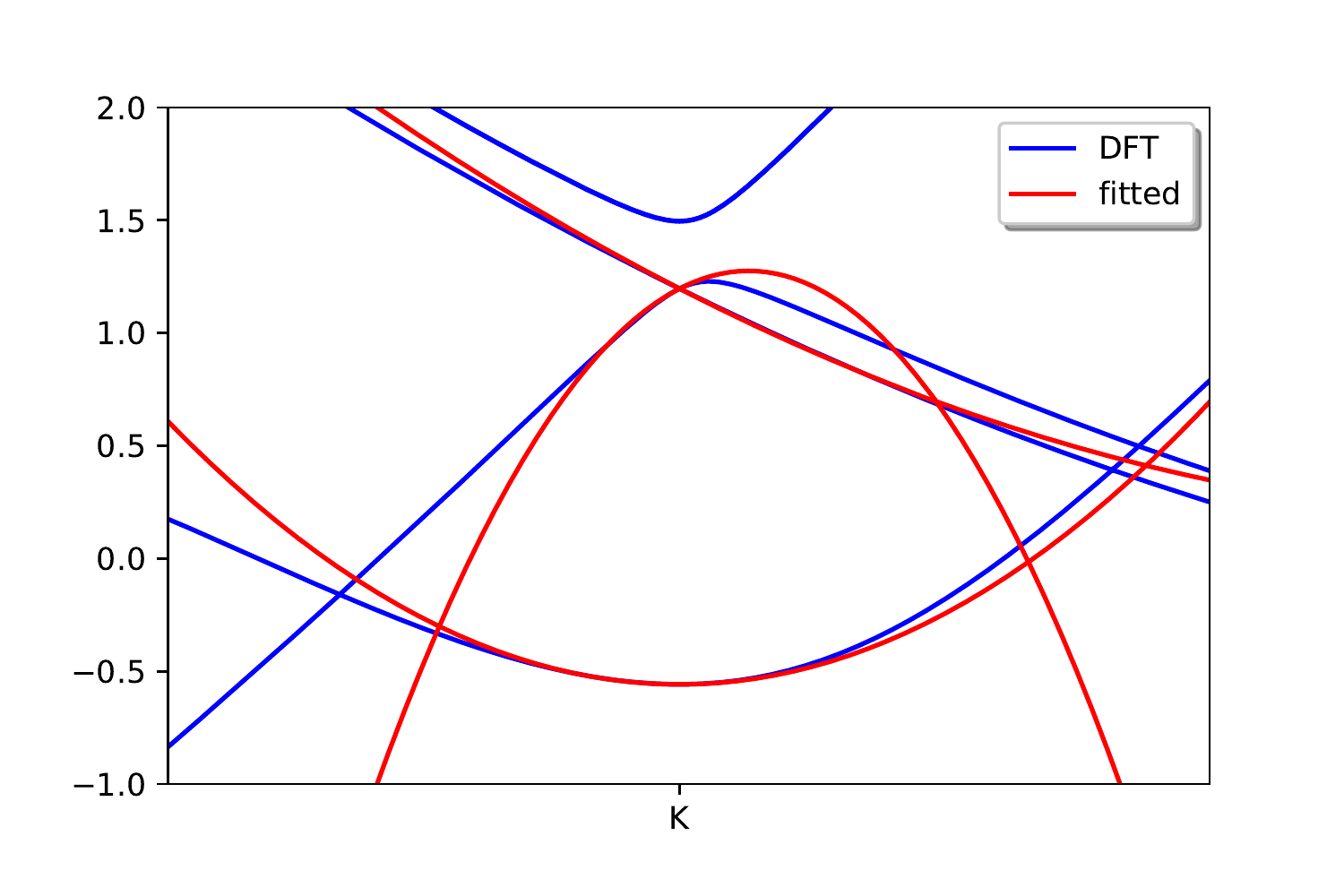}
  \caption{
           Comparison between DFT bands and the bands from the
continuum model near the K point of Al$_2$B$_2$. 
          }
  \label{fig:Al2B2-fit-K}
\end{figure}

%///////////////////////////////////////////////////////////////////////////////////////
\newpage
\subsection{Supporting information of the continuum model for Al$_2$B$_2$ configuration around the $\Gamma$ point.\label{SN2}}
\begin{table}[h!]
  \caption[]{\label{tab:Al2B2-G}
             Symmetrized matrices for the invariant
             expansion of the diagonal blocks $\mathcal{H}_{ii}$ for the point group $D_{6h}$.
            }
$\arraycolsep 0.8em
\renewcommand{\arraystretch}{1.2}
\begin{array}{clclcl} \hline \hline
\mbox{Block} &
\multicolumn{1}{l}{\mbox{Representations}} &
\multicolumn{2}{l}{\mbox{Symmetrized matrices}} &
\multicolumn{2}{l}{\mbox{Tensor components}} \\ \hline

\mathcal{H}_{11} & \Gamma_{6}^{+} \otimes \Gamma_6^{+ \ast}     & \Gamma_1^{+}: & \openone           & \Gamma_1^{+}: & 1; \; {k_x^2 + k_y^2}\\
                 & = \Gamma_1^{+} + \Gamma_2^{+} + \Gamma_6^{+} & \Gamma_2^{+}: & \sigma_z           & \Gamma_2^{+}: & - \\
                 &                                              & \Gamma_6^{+}: & \sigma_x, \sigma_y & \Gamma_6^{+}: & (k_y^2- k_x^2, 2k_x k_y); \; (k_x^4-6k_x^2k_y^2+k_y^4, 4k_xk_y^3+4k_x^3k_y) \\[2ex]

\mathcal{H}_{22} & \Gamma_2^{-} \otimes \Gamma_2^{- \ast}       & \Gamma_1^{+}: & \openone           & \Gamma_1^{+}: & 1; \;{k_x^2 + k_y^2} \\
                 & = \Gamma_1^{+} \\ \hline \hline
\end{array}$
\end{table}

\begin{equation}
E_{1,2}(k_x,k_y)={E_0}+A ({k_x}^2+{k_y}^2)\pm \sqrt{B^2 ({k_x}^2+{k_y}^2)^2+C^2 ({k_x}^2+{k_y}^2)^4-2BC({k_x}^6 - 15{k_x}^4 {k_y}^2+15k_x^2k_y^4-k_y^6)},
\end{equation}
\begin{equation}
E_{3}(k_x,k_y)={E'_0}+A'( {k_x}^2+ {k_y}^2).
\end{equation}
The fitted parameters are $E_0=-0.37$ eV, $A=-43.88$ eV\AA$^2$, $B=11.10$ eV\AA$^2$,$C=100.17$eV\AA$^4$, $E'_0=-2.85$ eV, $A'=12.36$ eV\AA$^2$ .
The fitted bands are shown in Supplementary Fig.~\ref{fig:Al2B2_fit_G}.
\begin{figure}[h!]
  \includegraphics[width=0.50\columnwidth]{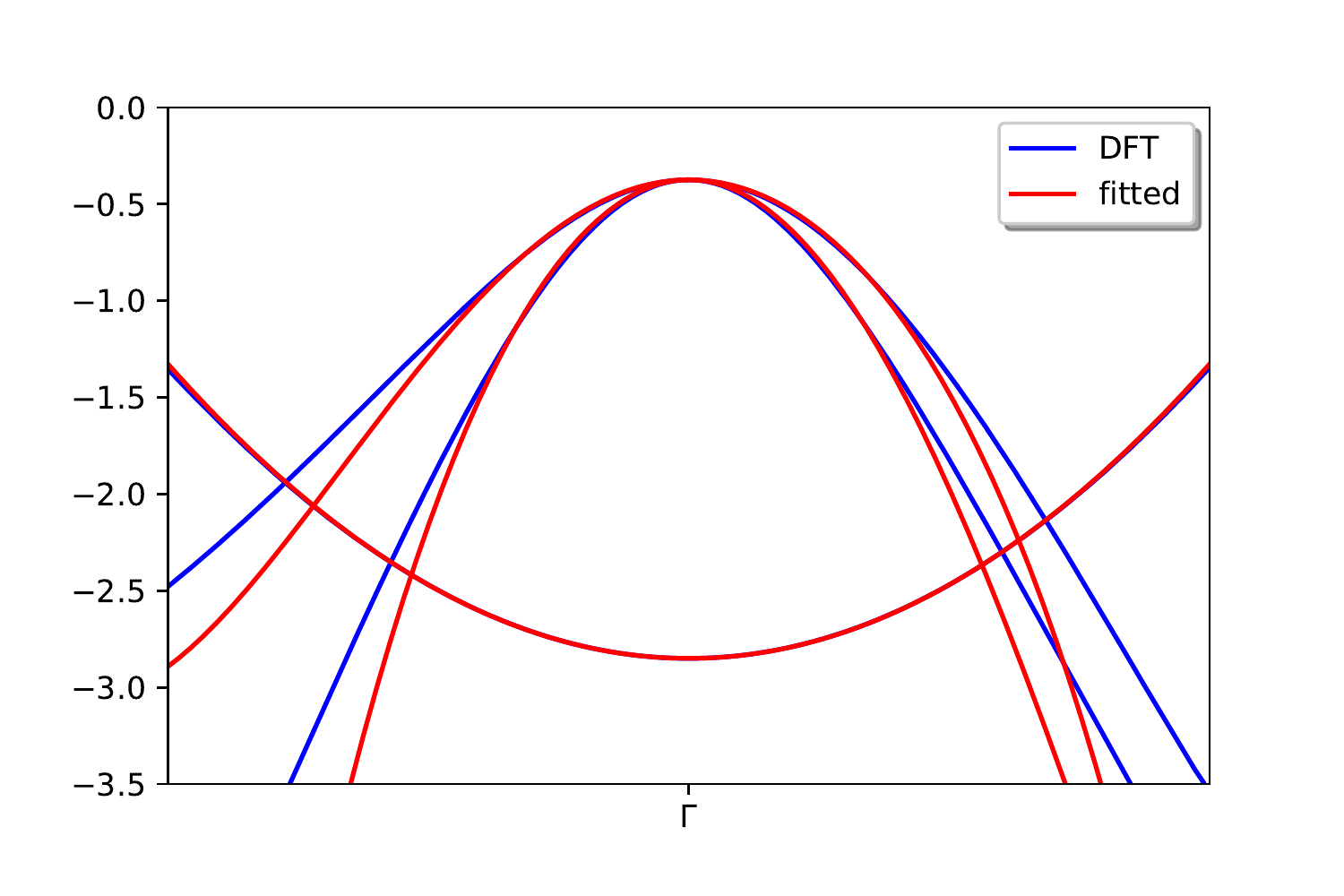}
  \caption{
                    Comparison between DFT bands and the bands from the
continuum model near the $\Gamma$ point of Al$_2$B$_2$. 
          }
  \label{fig:Al2B2_fit_G}
\end{figure}

%///////////////////////////////////////////////////////////////////////////////////////
\newpage
\subsection{Supporting information of the continuum model for AlB$_4$ configuration around the $\Gamma$ point.\label{SN3}}
\begin{table}[h!]
  \caption[]{\label{tab:AlB4-fit-G} 
             Symmetrized matrices for the invariant
             expansion of the diagonal blocks $\mathcal{H}_{ii}$ for the point group $D_{6h}$.
            }
$\arraycolsep 0.8em
\renewcommand{\arraystretch}{1.2}
\begin{array}{clclcl} \hline \hline
\mbox{Block} &
\multicolumn{1}{l}{\mbox{Representations}} &
\multicolumn{2}{l}{\mbox{Symmetrized matrices}} &
\multicolumn{2}{l}{\mbox{Tensor components}} \\ \hline

\mathcal{H}_{11} & \Gamma_{6}^{-} \otimes \Gamma_6^{- \ast}     & \Gamma_1^{+}: & \openone           & \Gamma_1^{+}: & 1; \; {k_x^2 + k_y^2}\\
                 & = \Gamma_1^{+} + \Gamma_2^{+} + \Gamma_6^{+} & \Gamma_2^{+}: & \sigma_z           & \Gamma_2^{+}: & - \\
                 &                                              & \Gamma_6^{+}: & \sigma_x, \sigma_y & \Gamma_6^{+}: & (k_y^2- k_x^2, 2k_x k_y); \; (k_x^4-6k_x^2k_y^2+k_y^4, 4k_xk_y^3+4k_x^3k_y) \\[2ex]

\mathcal{H}_{22} & \Gamma_{6}^{+} \otimes \Gamma_6^{+ \ast}     & \Gamma_1^{+}: & \openone           & \Gamma_1^{+}: & 1; \; {k_x^2 + k_y^2}\\
                 & = \Gamma_1^{+} + \Gamma_2^{+} + \Gamma_6^{+} & \Gamma_2^{+}: & \sigma_z           & \Gamma_2^{+}: & - \\
                 &                                              & \Gamma_6^{+}: & \sigma_x, \sigma_y & \Gamma_6^{+}: & (k_y^2- k_x^2, 2k_x k_y); \; (k_x^4-6k_x^2k_y^2+k_y^4, 4k_xk_y^3+4k_x^3k_y) \\[2ex]

\mathcal{H}_{33} & \Gamma_2^{-} \otimes \Gamma_2^{- \ast}       & \Gamma_1^{+}: & \openone           & \Gamma_1^{+}: & 1; \;{k_x^2 + k_y^2} \\
                 & = \Gamma_1^{+} \\ \hline \hline
\end{array}$
\end{table}

\begin{equation}
E_{1,2}(k_x,k_y)={E_0}+A ({k_x}^2+{k_y}^2)\pm \sqrt{B^2 ({k_x}^2+{k_y}^2)^2+C^2 ({k_x}^2+{k_y}^2)^4-2BC({k_x}^6 - 15{k_x}^4 {k_y}^2+15k_x^2k_y^4-k_y^6)},
\end{equation}
\begin{equation}
E_{3,4}(k_x,k_y)={E'_0}+A' ({k_x}^2+{k_y}^2)\pm \sqrt{B'^2 ({k_x}^2+{k_y}^2)^2+C'^2 ({k_x}^2+{k_y}^2)^4-2B'C'({k_x}^6 - 15{k_x}^4 {k_y}^2+15k_x^2k_y^4-k_y^6)},
\end{equation}
\begin{equation}
E_{5}(k_x,k_y)={E''_0}+A''( {k_x}^2+ {k_y}^2).
\end{equation}
The fitted parameters are
$E_0=0.28$ eV, $A=-42.10$ eV\AA$^2$, $B=12.55$ eV\AA$^2$,$C=74.19$eV\AA$^4$,
$E'_0=0.88$ eV, $A'=-36.14$ eV\AA$^2$, $B'=8.80$ eV\AA$^2$,$C'=58.35$eV\AA$^4$,
$E''_0=-2.72$ eV, $A''=11.64$ eV\AA$^2$ .
The fitted bands are shown in Supplementary Fig.~\ref{fig:AlB4-fit-G}.
\begin{figure}[h!]
  \includegraphics[width=0.50\columnwidth]{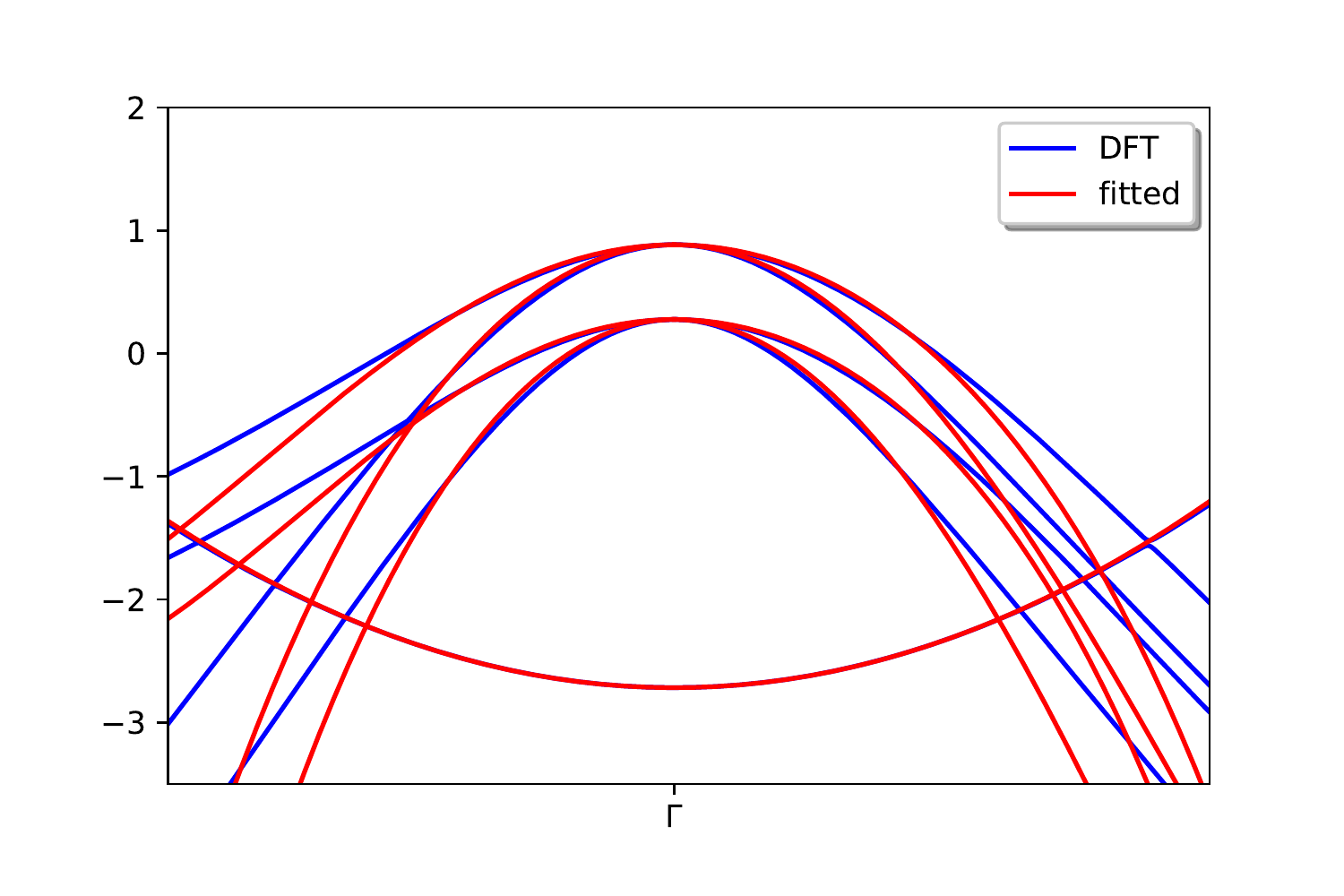}
  \caption{
           Comparison between DFT bands and the bands from the
continuum model near the $\Gamma$ point of AlB$_4$. 
          }
  \label{fig:AlB4-fit-G}
\end{figure}

%///////////////////////////////////////////////////////////////////////////////////////
\newpage
\subsection{Electronic properties of monolayer AlB$_2$.\label{SN4}}
\begin{figure}[h!]
  \includegraphics[width=0.50\columnwidth]{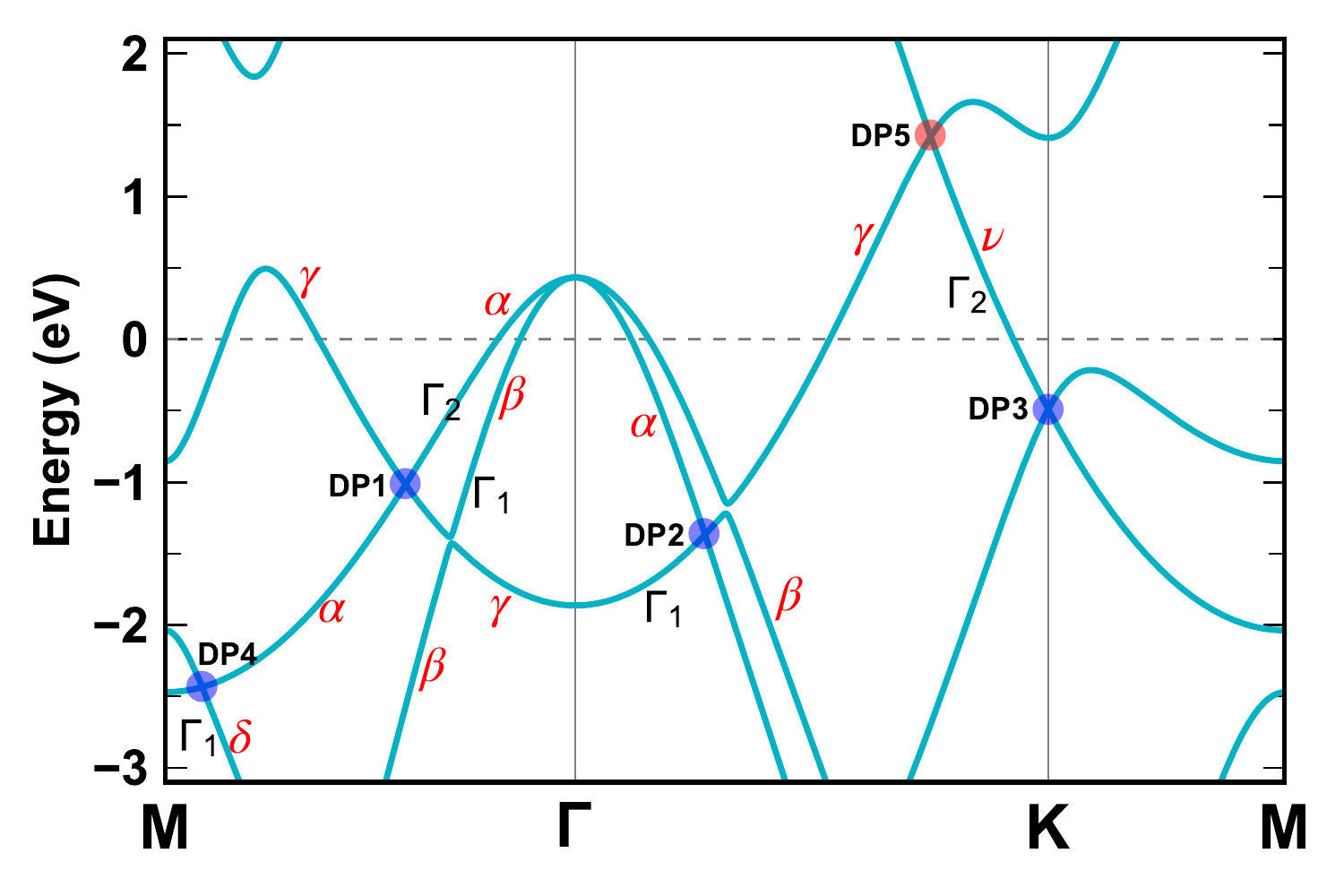}
  \caption{
          The electronic band structure of monolayer AlB$_2$ along the high symmetry
          paths M-$\Gamma$-K-M. The $p-$ and  $n-$type Dirac cones
          are shown by red and blue circles, respectively.
          }
  \label{fig:AlB2-BS}
\end{figure}

\begin{figure}[h!]
  \includegraphics[width=0.55\columnwidth]{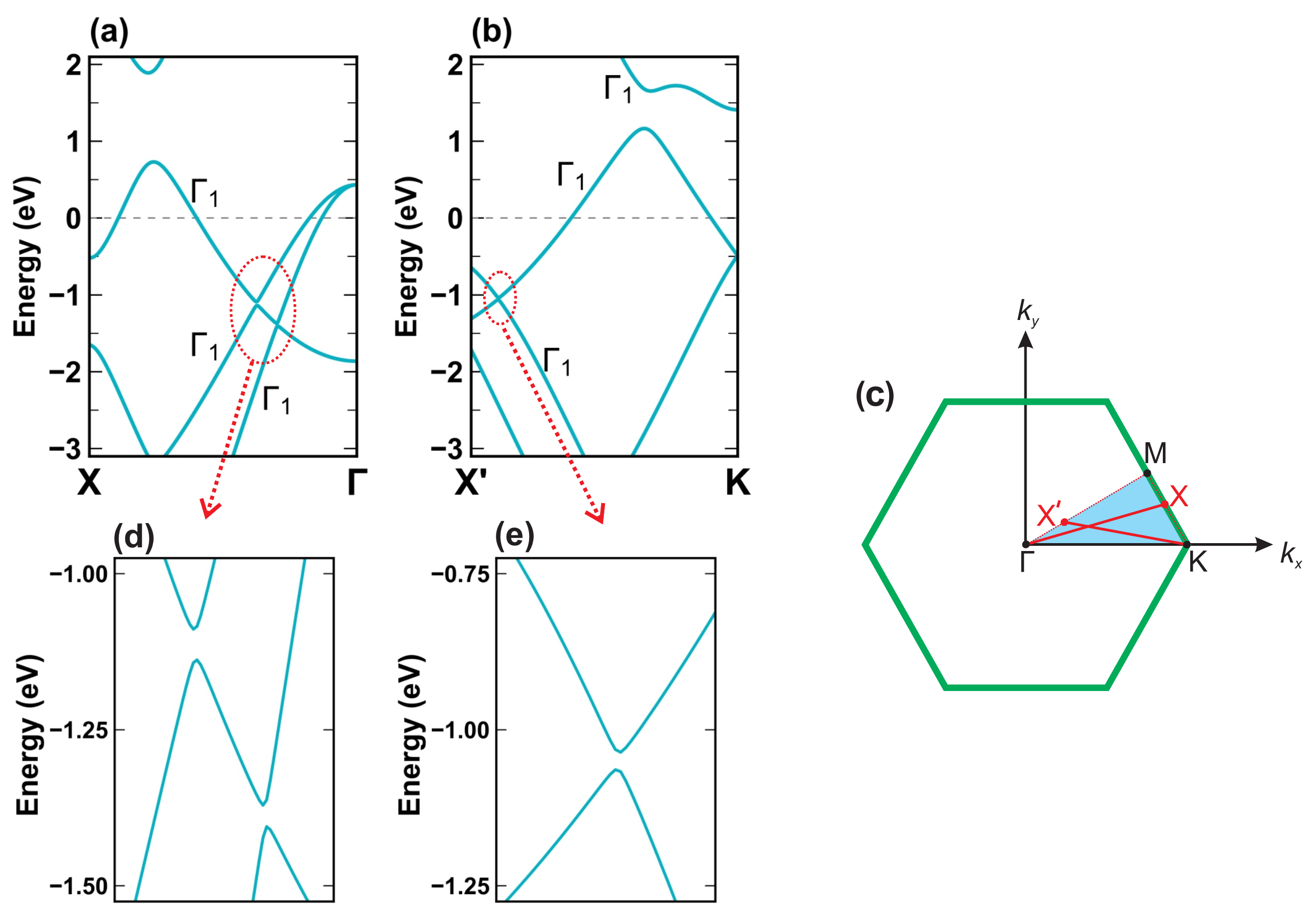}
  \caption{
            The electronic band structure of monolayer AlB$_2$
            along typical low-symmetry directions (a) X-$\Gamma$ and  (b) X$'$-K 
            as shown in (c). The zoomed-in band structures in the red ellipses are shown in (d) and (e).
          }
  \label{fig:AlB2-BS-XG-XK}
\end{figure}

The electronic band structure of freestanding monolayer AlB$_2$ is shown in
Fig.~\ref{fig:AlB2-BS}.  Within 2.5 eV of the Fermi level there are several bands  which cross to each other
to form  four $p-$type Dirac cones below (red circles) and a gapless $n-$type Dirac cone above (blue circle)
the Fermi surface, respectively.
We mark these bands by Greek letters as shown in the Figure.
Group theory analysis of these energy bands
shows that along  the $\Gamma$-M  and K-M symmetry directions the small groups are $C_s$ which 
has an out of plane symmetry element. Along these directions $\beta$, $\gamma$ and $\delta$ bands 
belong to the $\Gamma_1$ IR  which have positive mirror parity. Also, the IR of $\alpha$  and $\nu$ bands 
are $\Gamma_2$ with a negative mirror eigenvalue~\cite{altmann1994point}.   As a result,  
within the mentioned energy window one can observe 
the formation of four Dirac points DP1-DP4 due to the crossing of bands with opposite mirror parities. 
The remained Dirac point (DP3) is located at the K point which 
arises from the boron hexagonal lattice as the Dirac's cones of graphene.
Note that due to the absence of in-plane
mirror symmetry in monolayer AlB$_2$, along arbitrary low-symmetry 
directions X-$\Gamma$ and X$'$-K 
one can see no crossing feature (see Figs.~\ref{fig:AlB2-BS-XG-XK}(a) and (b)). Therefore, we do not expect to observe symmetry protected
nodal lines in this structure.
The existence 
of the $p-$type Dirac Fermions in this configuration has been confirmed
by recent experimental ARPES measurements~\cite{geng2020} which provides an excellent platform for
designing new nanoelectronics.\\


\iffalse

\fi

\end{document}